\documentclass[%
 reprint,
 superscriptaddress,
 amsmath,amssymb,
 aps,
 prb,
]{revtex4-2}
  
\makeatletter
\def\@hangfrom@section#1#2#3{\@hangfrom{#1#2}#3}
\def\@hangfroms@section#1#2{#1#2}
\makeatother
\usepackage[dvipsnames]{xcolor}
\usepackage{physics}
\usepackage{graphicx}
\usepackage{dcolumn}
\usepackage{bm}
\usepackage{float}
\usepackage{booktabs}
\usepackage{placeins}

\usepackage{xspace}
\usepackage{caption}
\usepackage{subcaption}
\usepackage{nameref}
\usepackage{multirow}
\usepackage{hyperref}
\hypersetup{colorlinks=true,allcolors=gray}
\usepackage{titlesec}
\titleformat{\subsection}
  {\normalfont\slshape\filcenter}
  {\thesubsection}{1em}{}
\begin{document}
\preprint{Draft}

\title{Fine-Tuning Unifies Foundational Machine-learned Interatomic Potential Architectures at \textit{ab initio} Accuracy}

\author{Jonas Hänseroth}
\email{jonas.haenseroth@tu-ilmenau.de}
\affiliation{Theoretical Solid State Physics, Institute of Physics, Technische Universität Ilmenau, 98693 Ilmenau, Germany}
\author{Aaron Flötotto}
\affiliation{Theoretical Solid State Physics, Institute of Physics, Technische Universität Ilmenau, 98693 Ilmenau, Germany}
\author{Muhammad Nawaz Qaisrani}
\affiliation{Theoretical Solid State Physics, Institute of Physics, Technische Universität Ilmenau, 98693 Ilmenau, Germany}
\author{Christian Dreßler}
\affiliation{Theoretical Solid State Physics, Institute of Physics, Technische Universität Ilmenau, 98693 Ilmenau, Germany}
\date{\today}
\begin{abstract}

\noindent This work demonstrates that fine-tuning transforms foundational machine-learned interatomic potentials (MLIPs) to achieve consistent, near-\textit{ab initio} accuracy across diverse architectures. Benchmarking five leading MLIP frameworks (MACE, GRACE, SevenNet, MatterSim, and ORB) across seven chemically diverse compounds reveals that fine-tuning universally enhances force predictions by factors of 5-15 and improves energy accuracy by 2-4 orders of magnitude. The investigated models span both equivariant and invariant, as well as conservative and non-conservative, architectures. While general-purpose foundation models are robust, they exhibit architecture-dependent deviations from \textit{ab initio} reference data; fine-tuning eliminates these discrepancies, enabling quantitatively accurate predictions of atomistic and structural properties. Using datasets constructed from equidistantly sampled frames of short \textit{ab initio} molecular dynamics trajectories, fine-tuning reduces force errors by an order of magnitude and harmonizes performance across all architectures. These findings establish fine-tuning as a universal route to achieving system-specific predictive accuracy while preserving the computational efficiency of MLIPs. To promote widespread adoption, we introduce the \textit{aMACEing Toolkit}, which provides a unified and reproducible interface for fine-tuning workflows across multiple MLIP frameworks.

\end{abstract}

\maketitle

\section*{Introduction}

The computational exploration of materials and molecular systems has long been constrained by the fundamental trade-off between accuracy and efficiency. \textit{Ab initio} molecular dynamics (AIMD), based on density functional theory (DFT), provides chemical accuracy but limits accessible system sizes to a few hundreds of atoms and timescales to picoseconds due to prohibitive computational costs. \cite{marx2000ab,tuckerman2002ab,iftimie2005ab}
Empirical force-field-based molecular dynamics, while enabling simulations of millions of atoms over multiple nanoseconds, significantly lacks accuracy, transferability, and chemical fidelity beyond its parameterization domain.\cite{plimpton1995computational,sutmann2002classical,brooks2021classical} This accuracy-efficiency dilemma has fundamentally restricted the scope of problems addressable through atomistic simulation.

Machine learning interatomic potentials (MLIPs) have emerged as a powerful approach, bridging near-\textit{ab initio} accuracy with the computational efficiency approaching that of classical methods.\cite{kabylda2025molecular,poltavsky2025crash,pravsnikar2024machine,wang2020machine} Early neural network potentials and Gaussian approximation potentials demonstrated the feasibility of learning potential energy surfaces directly from quantum chemical data.\cite{behler2007,bartok2010,friederich2021} The subsequent adoption of graph neural networks, equivariant architectures, and symmetry-preserving representations has dramatically improved the accuracy and transferability of MLIPs across diverse chemical systems.\cite{thomas2018tensor,batzner20223,mace_1,unke2021,reiser2022,grace_1,drautz2019}

The recent development of foundation models for atomistic simulations represents a paradigm shift toward universal, pre-trained potentials capable of modeling nearly the entire periodic table.\cite{jacobs2025practical,mace_mp, orbv3, grace_2} These models, trained on massive datasets spanning millions of DFT calculations from repositories such as the Materials Project, Open Materials, and Alexandria databases, offer remarkable zero-shot capabilities across diverse chemical systems.\cite{alexandria, mp_1, mp_2, omat24, omol25} Notable examples include MACE-MPA-0, GRACE foundation models trained on several datasets, MatterSim's universal potentials, ORB's v3 foundation model, and SevenNet's multi-fidelity models.\cite{mace_mp, grace_2, mattersim, sevennet_2, orbv3} However, despite their broad applicability, foundation models often fail to capture system-specific properties without further optimization.\cite{grunert2025, hanseroth2025optimizing,flototto2025large,weiske2025statistics,chen2025high,radova2025fine,liu2025fine,kaur2025data} 

Fine-tuning, the process of adapting pre-trained foundation models using system-specific training data - has emerged as a critical technique for achieving quantitative accuracy in specialized applications. Recent studies have demonstrated the effectiveness of fine-tuning approaches across various domains.\cite{grunert2025,flototto2025large,hanseroth2025optimizing,weiske2025statistics} Transfer learning strategies enable efficient adaptation of foundation models with relatively small datasets, typically requiring orders of magnitude less training data than training from scratch while achieving comparable accuracy.

Despite growing recognition of fine-tuning's importance, several challenges limit its widespread adoption. First, each MLIP framework implements fine-tuning differently, with distinct procedures, hyperparameters, and data formats creating technical barriers for researchers. Second, systematic comparisons of fine-tuning effectiveness across different frameworks and chemical systems remain limited, making it difficult to establish best practices. Third, the relationship between foundation model performance and fine-tuned model accuracy, as well as the impact of different training strategies, requires comprehensive investigation.

In this work, we address these challenges through a systematic evaluation of foundation model fine-tuning across five leading MLIP frameworks: MACE, GRACE, SevenNet, MatterSim, and ORB.\cite{mace_1, grace_1, sevennet_1, mattersim, orbv2} We investigate fine-tuning performance on seven diverse chemical systems: excellent solid state proton conductors such as cesium dihydrogen phosphate (CsH\textsubscript{2}PO\textsubscript{4}), and its derivative (Cs\textsubscript{7}(H\textsubscript{4}PO\textsubscript{4})(H\textsubscript{2}PO\textsubscript{4})\textsubscript{8}) 
containing the unusual tetrahydroxyphosphonium cation H\textsubscript{4}PO\textsubscript{4}\textsuperscript{+} , L-pyroglutamate-ammonium an organic crystal, that contains low barrier hydrogen bonds and exhibit non-aromatic intrinsic fluorescence when excited by near UV light, solvated phenol, aqueous potassium hydroxide solution, crystalline lithium silicide Li\textsubscript{13}Si\textsubscript{4}, and a molybdenum disulfide (MoS\textsubscript{2}) structure containing sulfur vacancies.\cite{qaisrani2025acid, miron2023carbonyl, stephens2021short} These systems were selected to span different chemical environments, bonding types, and dynamical phenomena relevant to contemporary materials research.

Our comprehensive analysis reveals that fine-tuning consistently and dramatically improves model accuracy across all frameworks and systems, with force errors typically decreasing by 5-15x and energy errors by 2-4 orders of magnitude. More importantly, we demonstrate that fine-tuning enables accurate reproduction of system-specific physical properties including diffusion coefficients, hydrogen bond dynamics, and structural correlations, that foundation models fail to capture. Through systematic comparison of training times, hyperparameter requirements, and final accuracies, we provide practical guidance for selecting appropriate frameworks and strategies for different applications.

To facilitate broader adoption of these methods, we introduce the aMACEing Toolkit, which provides a unified command-line interface for fine-tuning workflows across all supported MLIP frameworks. The toolkit streamlines the process by taking care of framework-specific complexities (such as training data formatting, training setup, interference with simulation environments, model conversion, performance evaluation and documentation of the computed investigation) while still providing access to advanced features, enabling researchers to focus on scientific questions rather than implementation details. Combined with comprehensive analysis capabilities for trajectory post-processing, the toolkit significantly lowers the barrier to utilizing state-of-the-art machine learning potentials in molecular dynamics research.

\section*{Methods}

\subsection*{Foundation Models and MLIP Frameworks}

We evaluate five prominent MLIP frameworks, all based on graph neural networks, each offering foundation models trained on comprehensive quantum chemical datasets. MACE employs higher-body-order equivariant message passing.\cite{mace_1, mace_2, mace_mp} GRACE utilizes graph extensions to the atomic cluster expansion.\cite{grace_1, grace_2} MatterSim is a invariant graph neural network based on the M3GNet architecture.\cite{mattersim,m3gnet} SevenNet offers scalable equivariant architectures with GPU-parallelism support and is based on the NequIP architecture.\cite{sevennet_1, sevennet_2, batzner20223} ORB is non-conservative and invariant, like MD-ET framework, directly predicting forces instead of computing the gradient of an energy function.\cite{orbv2, orbv3, eissler2025simple}

With the exception of MatterSim, all frameworks feature foundation models trained on combinations or subsets of the following databases: Materials Project, Alexandria Database, Open Materials 2024, and Open Molecules 2025.\cite{mp_1, mp_2, alexandria, omat24, omol25} The Microsoft Research AI for Science Team has trained foundation models with DFT-calculated data including a temperature range of 0-5000 K and pressure range of 0-1000 GPa.\cite{mattersim} This database is not publicly available. The Materials Project includes DFT calculations of over 200,000 materials.\cite{mp_1, mp_2} For training, the database is usually subsampled using pymatgen's StructureMatcher, resulting in a dataset containing 146,000 materials and 1.5 million DFT calculations (PBE+U), referred to as MPtrj.\cite{pymatgen,mptrj} The Alexandria database is composed of DFT structure relaxation trajectories of 3 million materials with 30 million DFT calculations (PBE+U), and for training, a sub-sampled dataset called sAlex is often used, including 10 million DFT calculations.\cite{alexandria, omat24} The Open Materials 2024 and Open Molecules 2025 datasets from Meta's FAIRchem each contain over 100 million DFT calculations (OMat24: PBE+U and OMol25: $\omega$B97M-V).\cite{omat24, omol25}

All these frameworks with their respective foundation models are ranked by Matbench Discovery and MLIP Arena as among the best-performing MLIPs currently available.\cite{matbench,chiang2025mlip}

\subsection*{Chemical Systems and Fine-Tuning Data Generation}

Our evaluation encompasses seven chemically diverse systems selected to represent different classes of materials and dynamical phenomena. CsH\textsubscript{2}PO\textsubscript{4} (CDP, 512 atoms, cubic unit cell, a=19.82 Å) serves as a model solid acid electrolyte exhibiting proton conductivity enabled by a strong as well as fluctuating hydrogen bond network.\cite{struct_cdp, boysenand2003high, dressler2020effect} Cs\textsubscript{7}(H\textsubscript{4}PO\textsubscript{4})(H\textsubscript{2}PO\textsubscript{4})\textsubscript{8} (CPP, 576 atoms, cubic, a=20.20 Å) represents a complex ionic solid with coexisting cationic and anionic phosphate groups.\cite{struct_cpp, dressler2023coexistence} L-pyroglutamate-ammonium (144 atoms, orthorhombic, a=5.15 Å, b=14.56 Å, c=17.05 Å) exemplifies organic molecular crystals with short hydrogen bonds. The phenol-water system (388 atoms, cubic, a=15.64 Å) models a simple organic molecule with a solvent.\cite{qaisrani2025acid, miron2023carbonyl, stephens2021short} Aqueous KOH solution (288 atoms, cubic, a=14.21 Å) represents electrolyte solutions with hydroxide ion transport.\cite{hanseroth2025optimizing, haenseroth2025ohlmc} Li\textsubscript{13}Si\textsubscript{4} (204 atoms, orthorhombic, a=15.90 Å, b=15.13 Å, c=13.40 Å) represents a lithium silicide with lithium ion diffusion, being a material of interest for battery research.\cite{kirsch2022atomistic,kirsch2025li+,zeilinger2013revision} Finally, the  memristive and two-dimensional 1H-MoS\textsubscript{2} (106 atoms, hexagonal, A=[19.15 Å, 0.0 Å, 0.0 Å], B=[9.58 Å, 16.59 Å, 0.0 Å], C=[0.0 Å, 0.0 Å, 40.0 Å]) with suflur vacancies exhibiting cooperative dynamics with high activation energies, completes our benchmark set.\cite{Li2018, Spetzler2024, flototto2025large}

For each system, training data consisting of 2000 configurations was extracted from Born-Oppenheimer AIMD trajectories computed using CP2K with BLYP or PBE exchange-correlation functional, Goedecker-Teter-Hutter pseudopotentials, and DZVP-MOLOPT basis sets.\cite{cp2k_1, cp2k_2, cp2k_3, cp2k_4, cp2k_5, cp2k_quickstep, cp2k_basis-set, cp2k_orb_trans, cp2k_gth-pseudopot1, cp2k_gth-pseudopot2, cp2k_gth-pseudopot3, blyp1, blyp2, pbe, nose1, nose2, nose3} Configurations were selected every 100th frame of the AIMD to span the main part of the relevant phase space at target temperatures with structural diversity representative of dynamical processes. Using this protocol, fine-tuning datasets consisting of positions, forces, and energies were computed for all systems.

\subsection*{Fine-Tuning Methodology}

\begin{figure}
\centering
\includegraphics[width=0.95\linewidth]{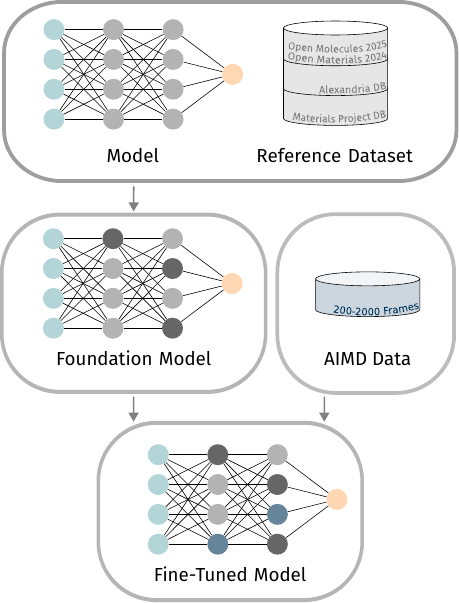}
\caption{Fine-tuning of a pre-trained foundation machine learning interatomic potential.}
\label{fig:finetuning}
\end{figure}

Fine-tuning protocols were implemented individually for each MLIP framework, with hyperparameters evaluated for each framework-system combination while maintaining consistency in training data and evaluation procedures. Training utilized 70-90\% of configurations for optimization, with remaining data reserved for validation and testing. To obtain fine-tuned foundation models capable of running stable MD simulations, key hyperparameter including learning rates (10\textsuperscript{-4}-10\textsuperscript{-2}), force-to-energy loss ratios (0.5-150), batch sizes (4 or 5), and epoch counts (200-2500) were adjusted to achieve MD-ready MLIPs for each system. Training was performed on GPU clusters with careful monitoring of convergence behavior. The fine-tuning protocol was applied to first-generation foundation models: MACE-MP-0, GRACE-1L-OAM, SevenNet-0, MatterSim Large, and ORB-v2 (see Figure \ref{fig:finetuning}).\cite{mace_mp,grace_2,sevennet_1,mattersim,orbv2} While the frameworks often offer more sophisticated foundation models that perform better on Matbench Discovery for a wide range of materials, fine-tuning these foundation models for specific systems with fewer parameters can achieve good performance while benefiting from the smaller model size, which can be used on hardware with less memory and run faster than more sophisticated models.\cite{matbench} For simplicity, the fine-tuning protocol was applied without incorporating active learning.

\subsection*{Evaluation Metrics and Analysis}

Model performance was assessed by recalculating first-principles structures excluded from the training set. In addition to force and energy mean absolute errors, models were evaluated on their ability to reproduce key physical properties derived from extended molecular dynamics simulations. Therefore, molecular dynamics simulations of 2-10 nanoseconds were performed using fine-tuned and foundation models: radial distribution functions characterizing structural correlations, mean square displacements and diffusion coefficients quantifying transport phenomena, and vector autocorrelation functions describing orientational dynamics (see Supporting Information Figures S1 - S19).

\subsection*{aMACEing Toolkit Implementation}

To facilitate reproducible fine-tuning workflows, we developed the aMACEing Toolkit, which provides unified interfaces for all supported MLIP frameworks. The toolkit handles data format conversions, generates framework-specific input files, manages job submission for high-performance computing environments, and provides comprehensive logging for reproducibility.

Key toolkit features include interactive question-and-answer interfaces for beginners, one-line command execution for automation, systematic benchmarking capabilities across multiple frameworks, built-in analysis tools for trajectory post-processing, and comprehensive documentation with practical examples. The toolkit can create input files for the Atomic Simulation Environment (ASE) and LAMMPS.\cite{ase,lammps} The modular architecture enables easy extension to additional frameworks while maintaining consistent user experiences.

\begin{figure}
\centering
\includegraphics[width=0.95\linewidth]{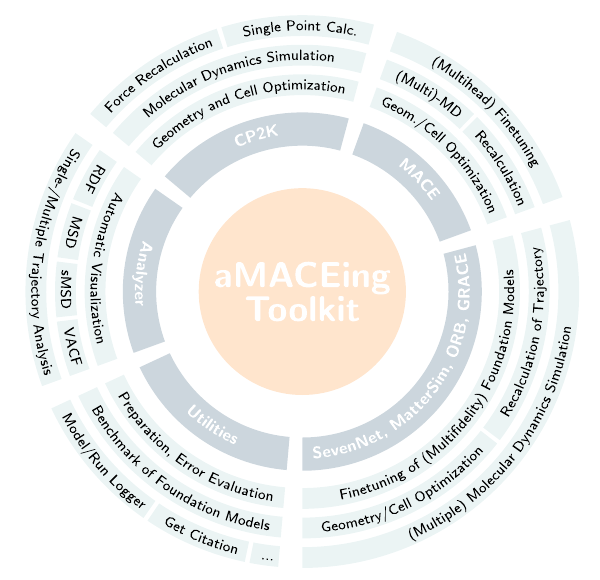}
\caption{Modules and functions of the aMACEing Toolkit.}
\label{fig:atk}
\end{figure}

Several other Python packages exist with somewhat similar functionalities, such as Janus-Core and AiiDA-TrainsPot.\cite{janus_core,aiida_trainspot} Janus-Core offers many modules for using multiple MLIPs to perform geometry optimizations, molecular dynamics, nudge elastic band calculations, and more. AiiDA-TrainsPot is a workflow that trains MLIPs automatically, currently only able to train MACE. The same limitation applies to fine-tuning with Janus-Core.

\section*{Results and Discussion}

\subsection*{Systematic Comparison of Foundation versus Fine-Tuned Models}

\begin{figure*}[]
    \centering
    \includegraphics[width=\textwidth]{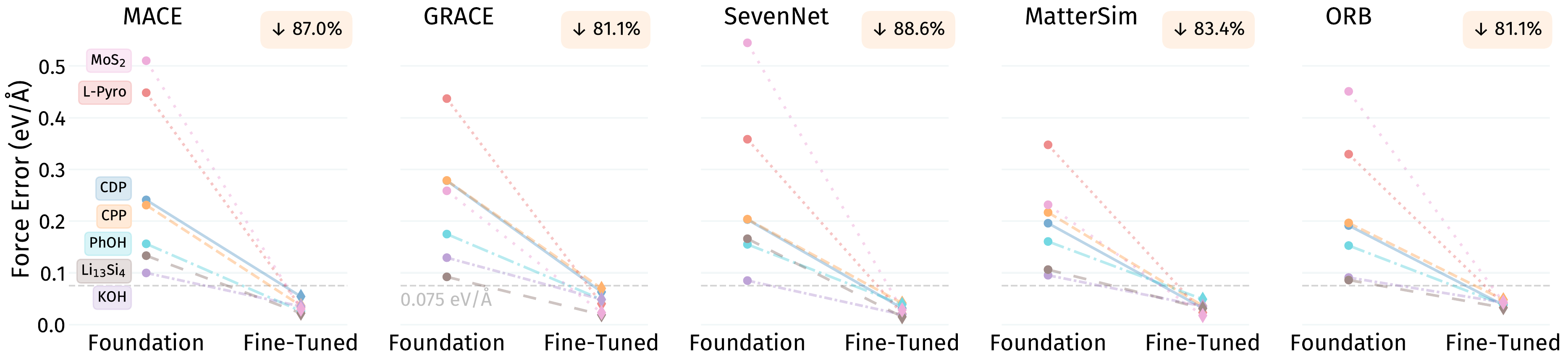}
    \caption{Root mean squared force errors for foundation and fine-tuned models across all evaluated systems: CsH\textsubscript{2}PO\textsubscript{4} (CDP), Cs\textsubscript{7}(H\textsubscript{4}PO\textsubscript{4})(H\textsubscript{2}PO\textsubscript{4})\textsubscript{8} (CPP), Li\textsubscript{13}Si\textsubscript{4}, solvated PhOH, aqueous KOH solution, L-pyroglutamate-ammonium (L-Pyro), MoS\textsubscript{2}; and frameworks with the respective foundation models: MACE-MP-0, GRACE-1L-OAM, SevenNet-0, MatterSim-Large, ORB-v2. Force errors in meV\,\AA$^{-1}$ and average error reduction in percent.} 
    \label{fig:frc_error}
\end{figure*}

Our comprehensive evaluation reveals dramatic and consistent improvements achieved through foundation model fine-tuning across all tested systems and frameworks. Figure~\ref{fig:frc_error} presents force prediction errors for foundation models versus their fine-tuned counterparts, demonstrating the universal effectiveness of this approach. Foundation models exhibit substantial errors ranging from 0.15-0.45 eV/Å for forces 
reflecting their general-purpose training on diverse chemical systems rather than optimization for specific applications. The numerical values including the energy error are listed in the Table S1 in the Supporting Information.

Fine-tuning consistently reduces these errors by remarkable margins. Force accuracy improves by factors of 5-15x, with mean absolute force errors decreasing to 0.02-0.07 eV/Å across all frameworks and systems. Energy errors also decrease substantially, often by several orders of magnitude, but their absolute values depend strongly on the underlying reference level (e.g., functional, basis set). Consequently, while the reduction in energy error highlights the overall consistency gained through fine-tuning, the improvements in force accuracy are the more physically meaningful indicator of enhanced model performance in molecular dynamics simulations. These improvements demonstrate that fine-tuning effectively adapts the broad knowledge encoded in foundation models to capture system-specific interactions with near-quantum chemical accuracy.

Notably, the magnitude of improvement shows limited dependence on the specific MLIP framework, suggesting that fine-tuning effectiveness is primarily determined by the quality and relevance of training data rather than architectural details. All frameworks: MACE, GRACE, SevenNet, MatterSim, and even the non-conservative framework ORB, achieve comparable final accuracies after fine-tuning, despite exhibiting different foundation model performance levels. These models were obtained without extensive hyperparameter optimization for every fine-tuning process; only small adjustments to the example values were needed for some systems. These findings have important practical implications, suggesting that framework selection might prioritize computational efficiency, training speed, or ease of use rather than foundation model accuracy alone. The most important step to achieve better accuracy is the fine-tuning step, as foundation models have not yet reached this level of precision. By using fine-tuned foundation models, a faster workflow requiring fewer computational resources is applied to obtain near \textit{ab initio} accurate trajectories of large systems on nanosecond length scales.

\subsection*{Training Efficiency and Computational Requirements}

Analysis of training times reveals significant variations across frameworks and systems, depending on system size, framework architecture, and hyperparameter choices. Table~\ref{tab:benchmark} presents a systematic comparison of the compute time for 100 epochs of fine-tuning for each framework and system, revealing framework-specific characteristics that influence practical deployment decisions.

GRACE generally exhibits the fastest training times, typically requiring less than one hour for 100 epochs of the systems studied, making it attractive for rapid prototyping and iterative refinement. MACE shows intermediate training times. SevenNet and MatterSim demonstrate variable performance depending on system characteristics, often requiring extended training periods. ORB demonstrates competitive training efficiency, particularly for system sizes where only computationally efficient non-conservative models like ORB are feasible.

\begin{table}
\centering
\caption{Computing time for 10,000 molecular dynamics steps and fine-tuning 100 epochs (2,000 data points) of a system containing 512 atoms on one NVIDIA A100.}
\begin{tabular}{lccc}
\toprule
Task & MACE & MACE+cueq & GRACE \\
\midrule
Molecular Dynamics  (s) & 390.2 & 383.5 & 312.6 \\
Model Fine-Tuning (min) & 134.0 & 51.8 & 40.9 \\
\midrule
\midrule
Task & SevenNet & MatterSim & ORB \\
\midrule
Molecular Dynamics (s) & 555.0 & 904.8 & 131.6 \\
Model Fine-Tuning (min) & 373.0 & 342.1 & 77.7 \\
\bottomrule
\end{tabular}
\label{tab:benchmark}
\end{table}

\subsection*{Physical Property Reproduction}

Beyond conventional energy and force accuracy metrics, we evaluate the ability of fine-tuned models to recover key physical properties, such as diffusion coefficients, radial distribution functions, and energy pathways, obtained from extended molecular dynamics simulations. This analysis reveals that fine-tuning not only improves agreement with reference forces and energies, but also enables accurate prediction of structural and dynamical observables that are often inaccessible to short-timescale \textit{ab initio} simulations and poorly captured by foundation models.

\begin{figure*}
\centering
\includegraphics[width=0.9\linewidth]{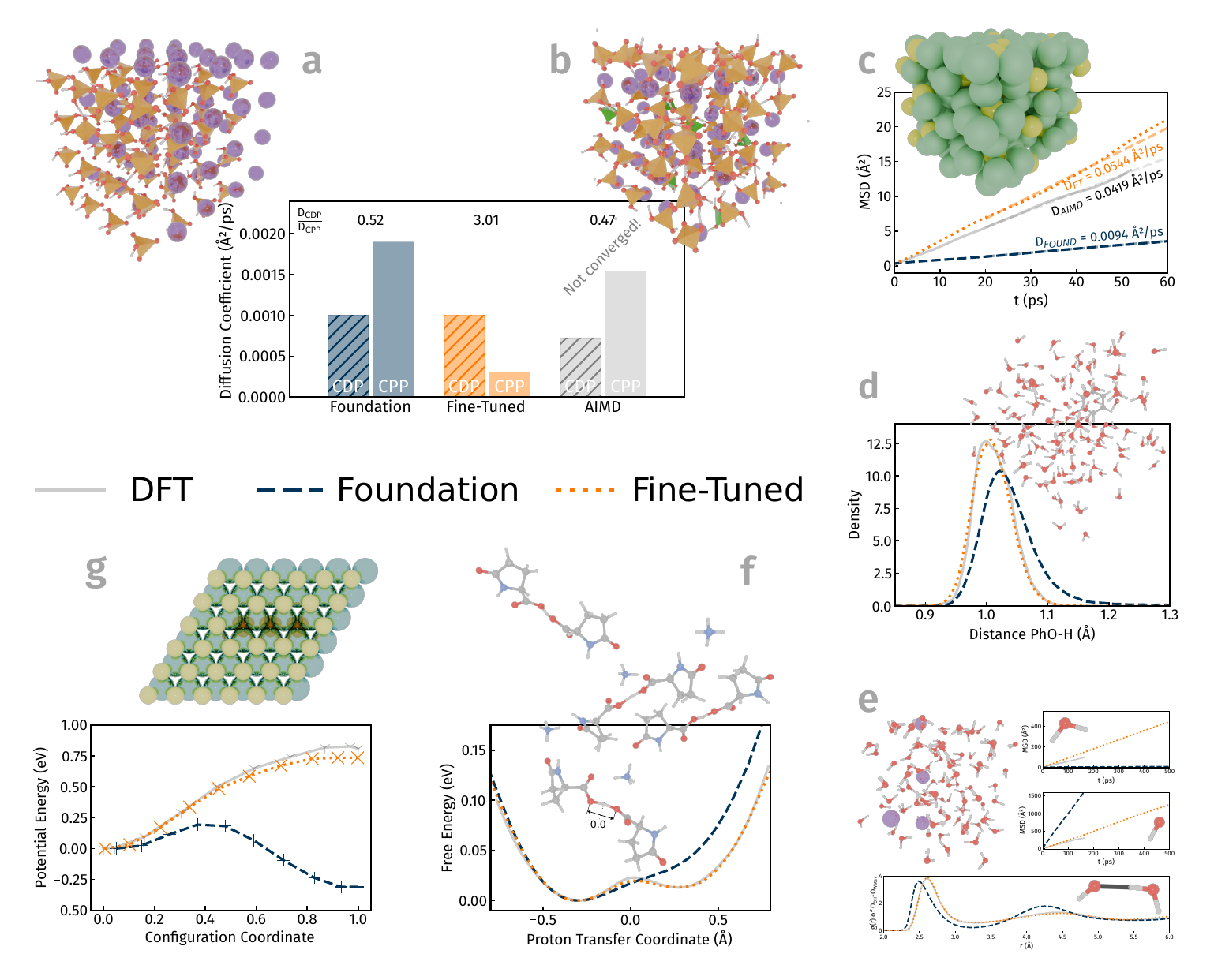}
\caption{Comparison of different physical properties obtained with first principles methods, foundation models and fine-tuned foundation models: (a \& b) CDP and CPP, proton diffusion coefficients ratios of D(CDP)/D(CPP) (MatterSim), (c) Li\textsubscript{13}Si\textsubscript{4}, lithium ion mean-squared displacements and diffusion coefficients (ORB), (d) Phenol in water, (O--H)\textsubscript{Hydroxyl-Group} bond length distribution (SevenNet), (e) KOH in water, water molecule and hydroxide ion mean-squared displacements and O\textsubscript{Hydroxide-Ion}-O\textsubscript{Water} radial distribution function (GRACE), (f) L-pyroglutamate-ammonium, free energy profiles along the proton transfer coordinate (ORB), (g) MoS\textsubscript{2}, potential energy curves for a sulfur jump into a neighboring line of sulfur vacancies (MACE).}
\label{fig:analysis_report}
\end{figure*}

The solid acids CsH\textsubscript{2}PO\textsubscript{4} and Cs\textsubscript{7}(H\textsubscript{4}PO\textsubscript{4})(H\textsubscript{2}PO\textsubscript{4})\textsubscript{8} (Figure~\ref{fig:analysis_report}a,b) are inorganic crystalline compounds that exhibit a superprotonic phase transition at elevated temperatures, accompanied by a drastic increase in proton conductivity. In the high-temperature phases of these compounds, the hydrogen-bond network becomes highly disordered, and the rotational dynamics of the anions approach those of a liquid state. This strong and dynamically fluctuating hydrogen-bond network enables efficient proton transfer through the Grotthuss mechanism. The overall proton diffusivity in these materials arises from a combination of the anion rotational rate and the proton transfer rate between neighboring anions. While proton transfer events within individual hydrogen bonds occur on the picosecond timescale, the rotational motion of the anions typically occurs on the order of several hundred picoseconds. Consequently, diffusion coefficients are challenging to converge in ab initio molecular dynamics simulations. Experimental studies indicate that proton diffusion is faster in CsH\textsubscript{2}PO\textsubscript{4} than in Cs\textsubscript{7}(H\textsubscript{4}PO\textsubscript{4})(H\textsubscript{2}PO\textsubscript{4})\textsubscript{8}.\cite{struct_cpp, wang2022phase} However, due to the limited timescales accessible to AIMD, even \textit{ab initio} simulations often fail to reproduce this qualitative difference in diffusion coefficients (Figure~\ref{fig:analysis_report}a,b).\cite{grunert2025, dressler2023coexistence} Similarly, many foundation models incorrectly predict the ratio of diffusion coefficients between the two compounds. In contrast, all fine-tuned foundation force fields correctly reproduced the experimental trend (see Supporting Information, Figures~S1-S7).

The diffusion coefficient for the lithium ions in the lithium silicide Li\textsubscript{13}Si\textsubscript{4} obtained with first principle methods is reproduced in the trajectories computed by the fine-tuned foundation model, while the foundation models consistently underestimates this value (see Figure \ref{fig:analysis_report}c and Supporting Information Figures S8 and S9). 

Given the critical role of the O--H stretch in determining phenol’s vibrational response and hydrogen-bonding behavior, the accuracy of various machine learning interatomic potentials in reproducing this structural feature was assessed. Figure \ref{fig:analysis_report}d presents the O--H distance distributions in phenol, showing that the fine-tuned models yield distributions closely aligned with the ab initio molecular dynamics reference, effectively capturing interactions with the surrounding solvent environment. In contrast, the foundation models produce broader and excessively delocalized distributions, reflecting an unrealistically soft potential along the O--H stretching coordinate. This artificial softening results in an overrepresentation of elongated O--H configurations, potentially biasing both infrared (IR) peak positions and intensities. Furthermore, the water structure surrounding the hydroxyl hydrogen of phenol is accurately reproduced by the fine-tuned model (see Supporting Information Figure S12).

Figure \ref{fig:analysis_report}e compares the mobilities of hydroxide ions and water molecules in aqueous potassium hydroxide solution. The foundation models fail to accurately reproduce the diffusion coefficients, underestimating or overestimating the diffusion of H\textsubscript{2}O, K\textsuperscript{+} and overestimating that of OH\textsuperscript{-} (see Supporting Information Figure S13 - S17). In contrast, the fine-tuned models show excellent agreement with the AIMD data. Moreover, the fine-tuned models more accurately captures the solvation environment of the hydroxide ion compared to the foundation models.

\begin{figure*}[]
    \centering
    \includegraphics[width=\textwidth]{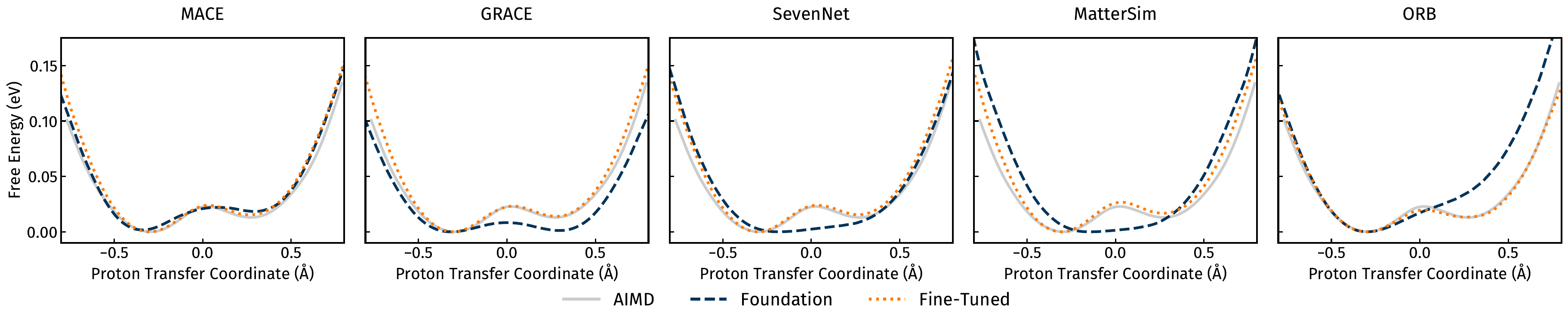}
    \caption{Free energy profiles along the proton transfer coordinate of the short-hydrogen-bond in L-pyroglutamate-NH\textsubscript{4} computed using different MLIP frameworks. Results from the foundation model and the fine-tuned foundation model are compared against AIMD reference data.} 
    \label{fig:lypro_profiles}
\end{figure*}

L-pyroglutamate-ammonium is another interesting system, an organic crystal that features a short hydrogen bond (SHB) with a donor-acceptor distance below 2.5\,\AA. In prior work, we showed that this SHB exhibit a low-barrier, asymmetric proton transfer (PT) potential in classical Born-Oppenheimer molecular dynamics (BOMD) simulations. The PT free energy barrier in these simulation is approximately 30\,meV. Although this barrier is shallow, it plays an important role in mediating the optical properties of this system.\cite{stephens2021short,miron2023carbonyl} When nuclear quantum effects are included via path-integral molecular dynamics, this barrier disappears and the SHB becomes symmetric and delocalized.\cite{qaisrani2025acid} However, the reference data for training machine-learned potentials in this work are derived from classical BOMD trajectories, which retain the asymmetric low-barrier profile. This distinction is crucial for evaluating ML model performance. As shown in Figure~\ref{fig:analysis_report}f, the correct classical reference profile is asymmetric with a shallow minimum. Most foundation models, however, fail to reproduce this structure. With the exception of MACE-MP-0, they instead predict flat or symmetric free energy surfaces, incorrectly mimicking quantum behavior that is absent from the training data. This results in inaccurate SHB dynamics and misleading structural interpretations. In contrast, all fine-tuned models across all frameworks recover the correct asymmetric profile and reproduce the low barrier observed in classical BOMD simulations (Figure~\ref{fig:lypro_profiles}). These findings demonstrate that subtle but chemically important features such as shallow PT barriers in SHBs are not captured by general-purpose models and require system-specific fine-tuning. 

In MoS\textsubscript{2}, the potential energy curves predicted by the foundation models substantially underestimate the vacancy jump barrier and exhibit an overall qualitatively incorrect trend. In contrast, the fine-tuned models accurately reproduce the DFT energy profile (see Figure \ref{fig:analysis_report}g).

A broader comparison is provided in the Supporting Information, where the performance of all investigated foundation and fine-tuned models is shown for every analysis presented here. Additional radial distribution function comparisons and other analyses are also listed there. These results demonstrate that fine-tuning enables not merely improved numerical accuracy but faithful reproduction of physical phenomena, making fine-tuned models suitable for quantitative prediction of experimentally observable properties. To illustrate this effect more concretely, a representative example of the exceptional performance achieved by fine-tuning is provided for the material L-pyroglutamate-ammonium in Figure~\ref{fig:lypro_profiles}. The free energy profiles predicted by the foundation models (despite MACE) deviate from the AIMD reference in a non-systematic manner. In contrast, fine-tuning substantially mitigates these discrepancies: all profiles obtained from molecular dynamics simulations with fine-tuned foundation models show excellent agreement with the AIMD reference data.

A comprehensive assessment across all investigated properties in the 70 multi-nanosecond MLIP simulations allows us to generalize the observations from Figure~\ref{fig:analysis_report} and the figures in the Supporting Information:
\begin{enumerate}
    \item The performance of the foundation models is noteworthy. In particular, these models are well suited for predicting non-dynamic properties in inorganic solids, such as radial distribution functions, where fine-tuning is sometimes unnecessary.
    \item For organic solids and general liquids, foundation models perform reasonably well but still show significant deviations from AIMD and experimental reference data.
    \item The differences in between the different foundation models are often substantial; none of the investigated models performs best in all cases, and the most accurate model is system-dependent.
    \item Fine-tuning systematically enhances force and energy predictions, yielding property predictions that are virtually indistinguishable from AIMD reference data across all MLIP frameworks. (Only one exception was identified: the potential energy curve for a sulfur jump in MoS\textsubscript{2} predicted with SevenNet.)
    \item In all cases, fine-tuning significantly reduces the spread in accuracy observed among foundation models for both property and force predictions.
\end{enumerate}

\subsection*{Framework Comparison and Recommendations}

All evaluated MLIP frameworks exhibit substantial performance improvements upon fine-tuning, while their corresponding foundation models already demonstrate remarkable versatility. With minimal fine-tuning effort, performed without active learning, all frameworks accurately reproduce first-principles trajectories and frequently achieve near-\textit{ab initio} precision. The fine-tuned models derived from conservative frameworks produce stable molecular dynamics simulations extending over multiple-nanosecond timescales for all investigated systems.
Overall, the differences between the frameworks are minor and do not lead to significant variations in their practical applicability.
Out of 35 fine-tuning attempts, only one, MoS\textsubscript{2} simulated with SevenNet, failed to reproduce physical properties with near-\textit{ab initio} accuracy. This finding underscores both the robustness of the evaluated approaches and the practical advantage of the aMACEing\_toolkit, which enables efficient testing and comparison of multiple MLIP frameworks, in contrast to other packages with limited model support. Nevertheless, subtle distinctions among the frameworks may still inform their selection for specific research objectives. MACE offers an excellent balance between training time, accuracy, and the availability of robust foundation models, making it particularly suitable for exploratory studies. GRACE combines outstanding accuracy with the fastest training and inference performance, enabling simulations over extended temporal and spatial scales. Through the integration of the new \texttt{cuEquivariance} package, which replaces the computational routines of the widely used equivariant neural network library \texttt{e3nn}, MACE achieves computation times comparable to GRACE, emerging as the most robust framework in our study.\cite{e3nn} ORB, owing to its non-conservative architecture, also delivers high computational speed; however, during extended molecular dynamics simulations, this same characteristic can sometimes cause instabilities that lead to the simulation box exploding. SevenNet and MatterSim achieve reliable accuracy, though their fine-tuning and inference stages are somewhat slower during molecular dynamics simulations. In summary, all investigated frameworks provide satisfactory accuracy and computational performance across the studied systems, indicating that the choice of MLIP framework for fine-tuning does not constitute a critical limiting factor in practice.

\section*{Conclusions}

This comprehensive evaluation demonstrates that fine-tuning foundation models represents a transformative approach for achieving near-\textit{ab initio} accuracy in specialized molecular dynamics applications. Our systematic study across five MLIP frameworks and diverse chemical systems establishes several key findings with broad implications for computational chemistry and materials science.

Fine-tuning consistently and dramatically improves model accuracy regardless of framework choice, with typical improvements of 5-15x for forces and 2-4 orders of magnitude for energies. This universality suggests that fine-tuning effectiveness depends primarily on training data quality and relevance rather than architectural details, providing flexibility in framework selection based on computational requirements and user preferences.

More importantly, fine-tuning enables accurate reproduction of system-specific physical properties that foundation models often fail to capture at this level of detail, including transport coefficients, structural correlations, and other dynamical phenomena. This capability transforms MLIPs from approximate simulation tools to predictive methods at near-\textit{ab initio} accuracy suitable for direct comparison with experimental measurements.

Given the observed independence of fine-tuning accuracy with respect to the underlying MLIP architecture and considering that no active learning protocol was employed for training data selection, we suggest that future community development efforts should prioritize inference speed, even at the cost of a minor loss in accuracy. 

The development of the aMACEing Toolkit addresses critical barriers limiting widespread adoption by providing unified workflows across multiple frameworks.
By abstracting technical complexities while maintaining flexibility, the toolkit enables researchers to leverage state-of-the-art methods without extensive specialized knowledge, potentially accelerating scientific discovery across diverse applications.

The universality of fine-tuning improvements across frameworks suggests that standardized benchmarking protocols and high-quality datasets could facilitate systematic comparison of different approaches. Such initiatives would benefit from the unified interfaces provided by tools like the aMACEing Toolkit, enabling large-scale collaborative evaluation studies.

Ultimately, this work establishes fine-tuning as an essential component of modern molecular simulation workflows, providing a practical pathway to near-quantum chemical accuracy for extended simulations. As foundation models continue to evolve and training datasets expand, fine-tuning approaches will likely become increasingly sophisticated, offering exciting opportunities for advancing our understanding of complex chemical systems across diverse applications in energy storage, catalysis, biological systems, and materials design.

\section*{Computational Details}

All fine-tuning calculations were performed using the respective MLIP framework implementations: MACE-torch 0.3.10, GRACE tensorpotential, SevenNet 0.11.2, MatterSim 1.1.2, and ORB 0.3.2 through their official APIs.\cite{mace_1,mace_2,grace_1,grace_2,sevennet_1,sevennet_2,mattersim,orbv2,orbv3} \textit{Ab initio} reference calculations were performed using CP2K 2025.1 with PBE and BLYP exchange-correlation functionals and GTH pseudopotentials.\cite{cp2k_1, cp2k_2, cp2k_3, cp2k_4, cp2k_5, cp2k_quickstep, cp2k_basis-set, cp2k_orb_trans, cp2k_gth-pseudopot1, cp2k_gth-pseudopot2, cp2k_gth-pseudopot3, blyp1, blyp2, pbe, nose1, nose2, nose3} Training data consisted of 2000 configurations per system extracted from AIMD trajectories at relevant temperatures (300-600 K depending on system). Molecular dynamics simulations for property evaluation were performed using LAMMPS and ASE with system-specific temperatures using Nosé-Hoover chain thermostats.\cite{lammps,ase, nose1, nose2, nose3} Calculations were performed on the compute cluster of Technische Universität Ilmenau using NVIDIA A100 GPUs for training and MD simulations.

The aMACEing Toolkit is available at \url{https://github.com/jhaens/amaceing_toolkit} with comprehensive documentation at \url{https://amaceing-toolkit.readthedocs.io}.

\section*{Acknowledgments}

We gratefully acknowledge funding from the Thüringer Aufbaubank (TAB) and the European Social Fund Plus (ESF+): KapMemLyse, grant no. 2024 FGR 0081 / 0082, the Carl-Zeiss-Stiftung through project SustEntMat and computational resources provided by the Compute Center of Technische Universität Ilmenau. We thank Henning Schwanbeck for technical support and system administration.

\section*{Data Availability}

The fine-tuning, evaluation, and production run workflows are available through the aMACEing Toolkit repository: \url{https://github.com/jhaens/amaceing_toolkit}. The complete production input, evaluation data, training datasets and the fine-tuned models are available at \url{https://doi.org/10.5281/zenodo.17438087}. The large trajectory data is upon request from the authors.

\FloatBarrier
\bibliography{bibliography.bib}
\end{document}


\FloatBarrier
\newpage

\begin{table}[ht]
\centering
\caption{Absolute force and energy errors for foundation and fine-tuned models across all evaluated systems. Errors: force (eV\,\AA$^{-1}$), energy per atom (eV).}
\label{tab:all_errors}
\resizebox{1\textwidth}{!}{
\begin{tabular}{llcccccccccc}
\toprule
System & Model & \multicolumn{2}{c}{MACE-MP-0} & \multicolumn{2}{c}{GRACE-1L-OAM} & \multicolumn{2}{c}{SevenNet-0} & \multicolumn{2}{c}{MatterSim-Large} & \multicolumn{2}{c}{ORB-v2} \\
\cmidrule(lr){3-4} \cmidrule(lr){5-6} \cmidrule(lr){7-8} \cmidrule(lr){9-10} \cmidrule(lr){11-12}
& &
Force & Energy & Force & Energy & Force & Energy & Force & Energy & Force & Energy \\
\midrule
\multirow{2}{*}{CsH\textsubscript{2}PO\textsubscript{4}}
 & Foundation & 0.2411 & 307.68 & 0.2782 & 307.69 & 0.2031 & 307.69 & 0.1960 & 307.69 & 0.1916 & 307.69 \\
 & Fine-tuned & 0.0543 & 0.00041 & 0.0631 & 0.00017 & 0.0316 & 0.00096 & 0.032 & 0.00050 & 0.0378 & 0.00192 \\
\midrule
\multirow{2}{*}{Cs\textsubscript{7}(H\textsubscript{4}PO\textsubscript{4})(H\textsubscript{2}PO\textsubscript{4})\textsubscript{8}}
 & Foundation & 0.2310 & 292.84 & 0.2787 & 292.85 & 0.2039 & 292.84 & 0.2169 & 292.85 & 0.1966 & 292.85 \\
 & Fine-tuned & 0.0364 & 0.00015 & 0.0699 & 0.00017 & 0.0414 & 0.00077 & 0.0363 & 0.00237 & 0.0480 & 0.00249 \\
\midrule
\multirow{2}{*}{L-pyroglutamate-NH\textsubscript{4}}
 & Foundation & 0.4484 & 139.42 & 0.4370 & 139.43 & 0.3584 & 139.44 & 0.3476 & 139.44 & 0.3295 & 139.43 \\
 & Fine-tuned & 0.0333 & 0.00153 & 0.0403 & 0.00013 & 0.0211 & 0.00183 & 0.0232 & 0.00320 & 0.0404 & 0.00075 \\
\midrule
\multirow{2}{*}{PhOH in H\textsubscript{2}O}
 & Foundation & 0.1562 & 149.64 & 0.1750 & 149.64 & 0.1551 & 149.64 & 0.1607 & 149.64 & 0.1528 & 149.63 \\
 & Fine-tuned & 0.0261 & 0.00052 & 0.0485 & 0.00022 & 0.0388 & 0.00312 & 0.0490 & 0.00940 & 0.0383 & 0.00150 \\
\midrule
\multirow{2}{*}{KOH in H\textsubscript{2}O}
 & Foundation & 0.0999 & 161.61 & 0.1294 & 161.62 & 0.0851 & 161.62 & 0.0954 & 161.62 & 0.0910 & 161.61 \\
 & Fine-tuned & 0.0351 & 0.00341 & 0.0485 & 0.00213 & 0.0193 & 0.00216 & 0.0354 & 0.00594 & 0.0426 & 0.00272 \\
\midrule
\multirow{2}{*}{Li\textsubscript{13}Si\textsubscript{4}}
 & Foundation & 0.1333 & 177.21 & 0.0923 & 177.23 & 0.1660 & 177.22 & 0.1063 & 177.23 & 0.0861 & 177.22 \\
 & Fine-tuned & 0.0220 & 0.00310 & 0.0190 & 0.00143 & 0.0151 & 0.00186 & 0.0313 & 0.00234 & 0.0327 & 0.00405 \\
\midrule
\multirow{2}{*}{MoS\textsubscript{2}}
 & Foundation & 0.5103 & 807.51 & 0.2587 & 807.52 & 0.5448 & 807.51 & 0.2317 & 807.53 & 0.4510 & 807.55 \\
 & Fine-tuned & 0.0299 & 0.00109 & 0.02299 & 0.00350 & 0.0284 & 0.00013 & 0.0175 & 0.00023 & 0.0431 & 0.00136 \\
\bottomrule
\end{tabular}}
\end{table}

\begin{table}
\centering
\caption{Computing time in minutes for fine-tuning across different MLIP frameworks and chemical systems per 100 epochs on one NVIDIA A100.}
\begin{tabular}{lccccc}
\toprule
System & MACE & GRACE & SevenNet & MatterSim & ORB \\
\midrule
CDP & 134.0 & 40.9 & 373.0 & 342.1 & 77.7 \\
CPP & 167.0 & 36.7 & 364.7 & 456.0 & 108.0 \\
L-PyroNH\textsubscript{4} & 37.5 & 12.2 & 188.5 & 128.0 & 44.0 \\
PhOH & 42.3 & 16.8 & 45.5 & 45.4 & 8.8 \\
KOH & 85.0 & 26.1 & 359.2 & 338.0 & 150.5 \\
Li\textsubscript{13}Si\textsubscript{4} & 67.0 & 21.0 & 164.0 & 178.6 & 58.7 \\
\bottomrule
\end{tabular}
\label{tab:traintime}
\end{table}

\begin{table}
\centering
\caption{Computing time for 10,000 molecular dynamics steps and fine-tuning 100 epochs (2,000 data points) of a system containing 512 atoms on one NVIDIA A100.}
\begin{tabular}{lccc}
\toprule
Task & MACE & MACE+cueq & GRACE \\
\midrule
MD foundation (s) & 412.6 & 385.1 & 292.2 \\
MD fine-tuned  (s) & 390.2 & 383.5 & 312.6 \\
Fine-tuning model (min) & 134.0 & 51.8 & 40.9 \\
\midrule
\midrule
Task & SevenNet & MatterSim & ORB \\
\midrule
MD foundation (s) & 549.0 & 915.6 & 131.6 \\
MD fine-tuned (s) & 555.0 & 904.8 & 131.6 \\
Fine-tuning model (min) & 373.0 & 342.1 & 77.7 \\
\bottomrule
\end{tabular}
\label{tab:benchmark}
\end{table}

\begin{table}
\centering
\caption{Hyperparameters used for fine-tuning across different MLIP frameworks and chemical systems. Learning rates, force weights, and epoch counts show both framework-specific preferences and system-dependent requirements.}
\begin{tabular}{lccccc}
\toprule
System & Framework & Learning Rate & Force Weight & Batch Size & Epochs \\
\midrule
\multirow{5}{*}{CsH\textsubscript{2}PO\textsubscript{4}}
 & MACE & 0.01 & 100 & 5 & 200 \\
 & GRACE & 0.002 & 150 & 4 & 2000 \\
 & SevenNet & 0.01 & 1 & 5 & 250 \\
 & MatterSim & 0.0005 & 0.5 & 5 & 500 \\
 & ORB & 0.0003 & 0.5 & 4 & 1650 \\
\midrule
\multirow{5}{*}{Cs\textsubscript{7}(H\textsubscript{4}PO\textsubscript{4})(H\textsubscript{2}PO\textsubscript{4})\textsubscript{8}}
 & MACE & 0.01 & 100 & 5 & 200 \\
 & GRACE & 0.002 & 50 & 4 & 2500 \\
 & SevenNet & 0.004 & 1 & 4 & 300 \\
 & MatterSim & 0.0005 & 0.5 & 5 & 500 \\
 & ORB & 0.0003 & 1 & 4 & 400 \\
\midrule
\multirow{5}{*}{L-pyroglutamate-NH\textsubscript{4}}
 & MACE & 0.01 & 10 & 5 & 200 \\
 & GRACE & 0.002 & 100 & 4 & 1000 \\
 & SevenNet & 0.01 & 100 & 4 & 200 \\
 & MatterSim & 0.0005 & 0.5 & 5 & 500 \\
 & ORB & 0.0003 & 0.5 & 4 & 400 \\
\midrule
\multirow{5}{*}{PhOH in H\textsubscript{2}O}
 & MACE & 0.01 & 10 & 5 & 200 \\
 & GRACE & 0.002 & 100 & 4 & 500 \\
 & SevenNet & 0.004 & 100 & 4 & 400 \\
 & MatterSim & 0.0005 & 0.25 & 5 & 500 \\
 & ORB & 0.0002 & 0.25 & 8 & 800 \\
\midrule
\multirow{5}{*}{KOH in H\textsubscript{2}O}
 & MACE & 0.01 & 100 & 5 & 200 \\
 & GRACE & 0.001 & 5 & 4 & 500 \\
 & SevenNet & 0.01 & 100 & 4 & 200 \\
 & MatterSim & 0.0005 & 0.5 & 5 & 500 \\
 & ORB & 0.0003 & 1 & 4 & 200 \\
\midrule
\multirow{5}{*}{Li\textsubscript{13}Si\textsubscript{4}}
 & MACE & 0.01 & 10 & 5 & 200 \\
 & GRACE & 0.002 & 100 & 4 & 500 \\
 & SevenNet & 0.004 & 50 & 4 & 200 \\
 & MatterSim & 0.0005 & 0.5 & 5 & 350 \\
 & ORB & 0.0003 & 0.75 & 4 & 1250 \\
\midrule
\multirow{5}{*}{MoS\textsubscript{2}}
 & MACE & 0.01 & 100 & 5 & 200 \\
 & GRACE & 0.001 & 100 & 4 & 1000 \\
 & SevenNet & 0.01 & 1 & 5 & 400 \\
 & MatterSim & 0.001 & 10 & 5 & 500 \\
 & ORB & 0.0003 & 0.5 & 4 & 750 \\
\bottomrule
\end{tabular}
\label{tab:hyperparameters}
\end{table}

\FloatBarrier
\newpage
\section{System A: CsH\textsubscript{2}PO\textsubscript{4}}

\begin{figure*}[]
    \centering
    \includegraphics[width=\textwidth]{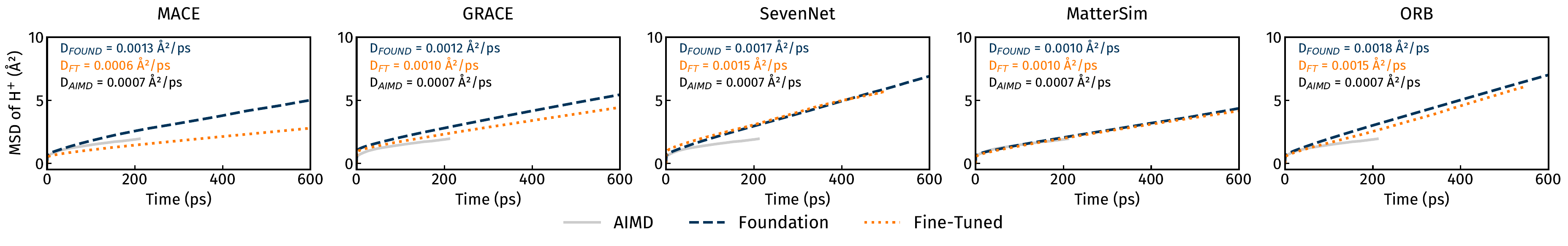}
    \caption{Mean-squared displacements of H\textsuperscript{+} in CsH\textsubscript{2}PO\textsubscript{4} computed using different MLIP frameworks. Results from the foundation model and the fine-tuned foundation model are compared against AIMD reference data.}
    \label{img:cdp_msd}
\end{figure*}

\begin{figure*}[]
    \centering
    \includegraphics[width=0.55\textwidth]{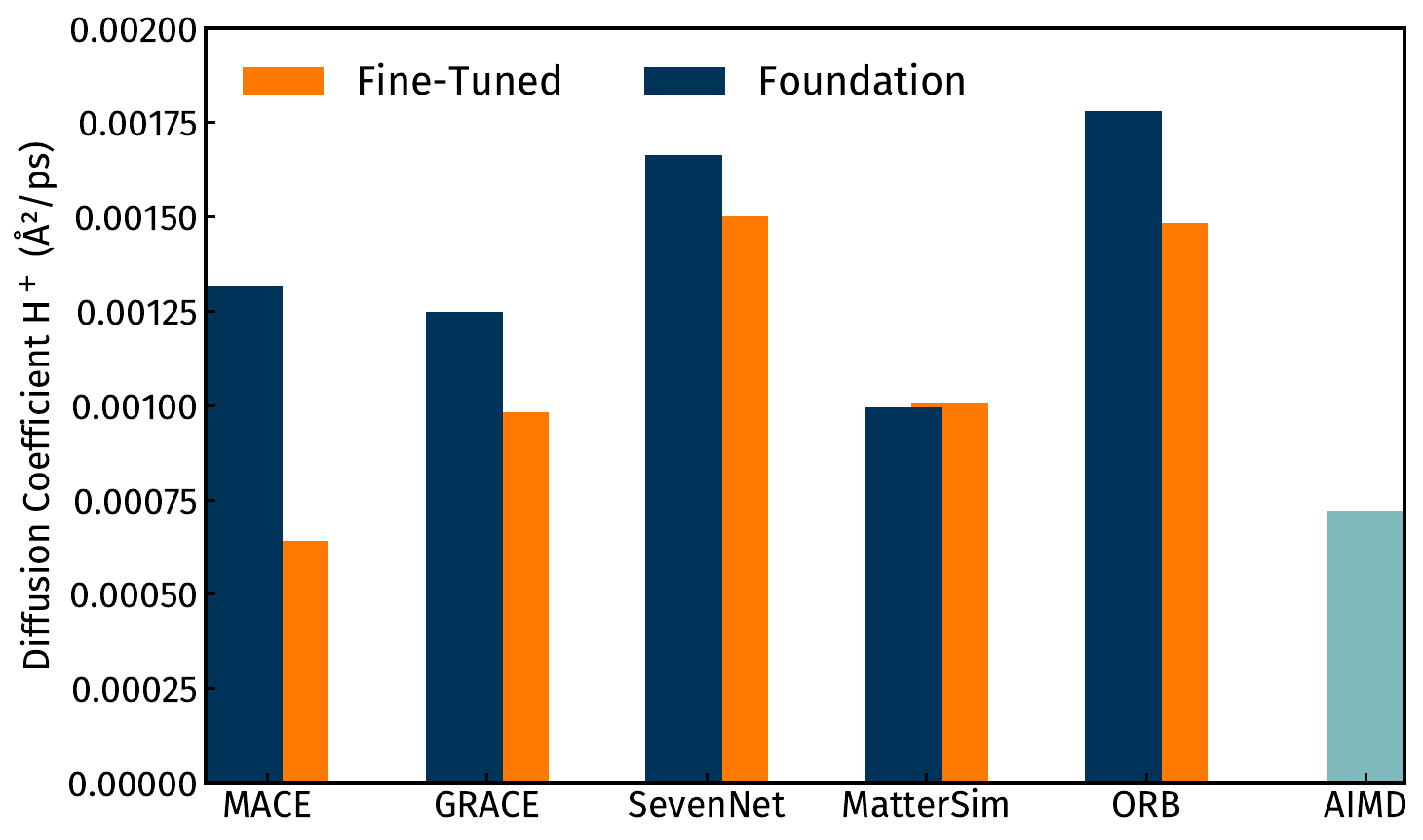}
    \caption{Diffusion coefficients of H\textsuperscript{+} in CsH\textsubscript{2}PO\textsubscript{4} computed using different MLIP frameworks from the mean-square displacements (see Figure \ref{img:cdp_msd}). Results from the foundation model and the fine-tuned foundation model are compared against AIMD reference data. }
    \label{img:cdp_diffcoeff}
\end{figure*}

\begin{figure*}[]
    \centering
    \includegraphics[width=\textwidth]{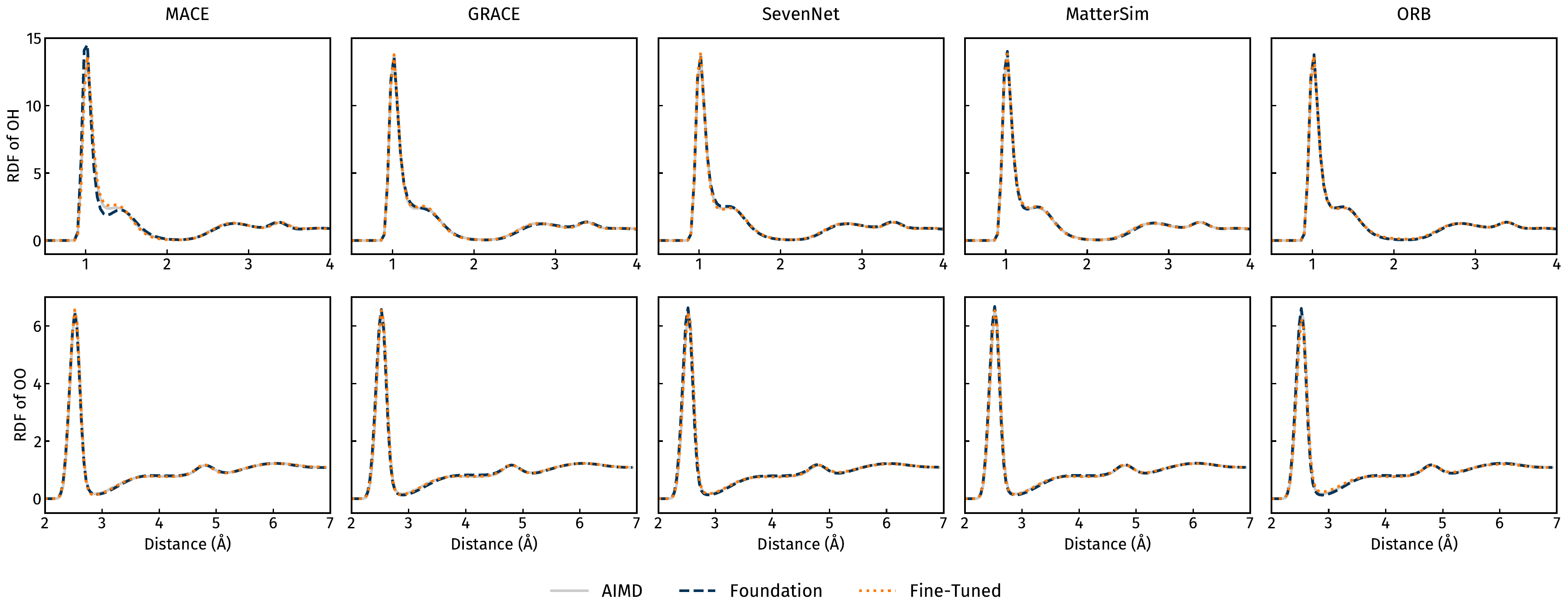}
    \caption{Radial-distribution functions of O-H and O-O in CsH\textsubscript{2}PO\textsubscript{4} computed using different MLIP frameworks. Results from the foundation model and the fine-tuned foundation model are compared against AIMD reference data.} 
    \label{img:cdp_rdf}
\end{figure*}

\FloatBarrier
\newpage
\section{System B: Cs\textsubscript{7}(H\textsubscript{4}PO\textsubscript{4})(H\textsubscript{2}PO\textsubscript{4})\textsubscript{8}}

\begin{figure*}[]
    \centering
    \includegraphics[width=\textwidth]{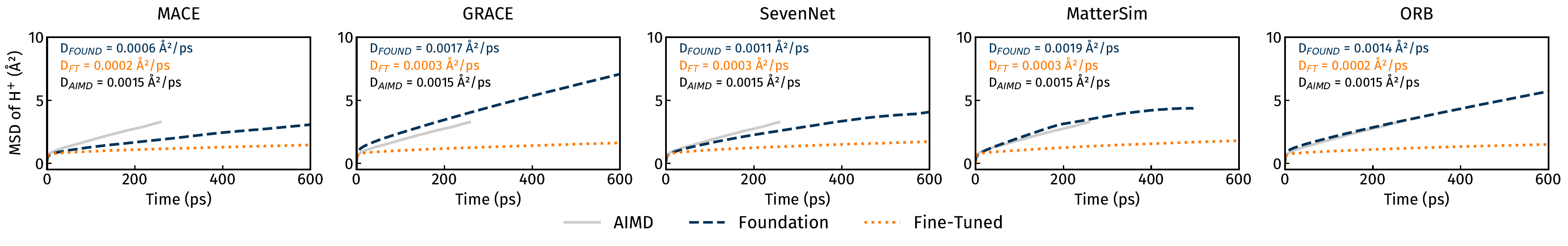}
    \caption{Mean-squared displacements of H\textsuperscript{+} in Cs\textsubscript{7}(H\textsubscript{4}PO\textsubscript{4})(H\textsubscript{2}PO\textsubscript{4})\textsubscript{8} computed using different MLIP frameworks. Results from the foundation model and the fine-tuned foundation model are compared against AIMD reference data.} 
    \label{img:cpp_msd}
\end{figure*}

\begin{figure*}[]
    \centering
    \includegraphics[width=0.55\textwidth]{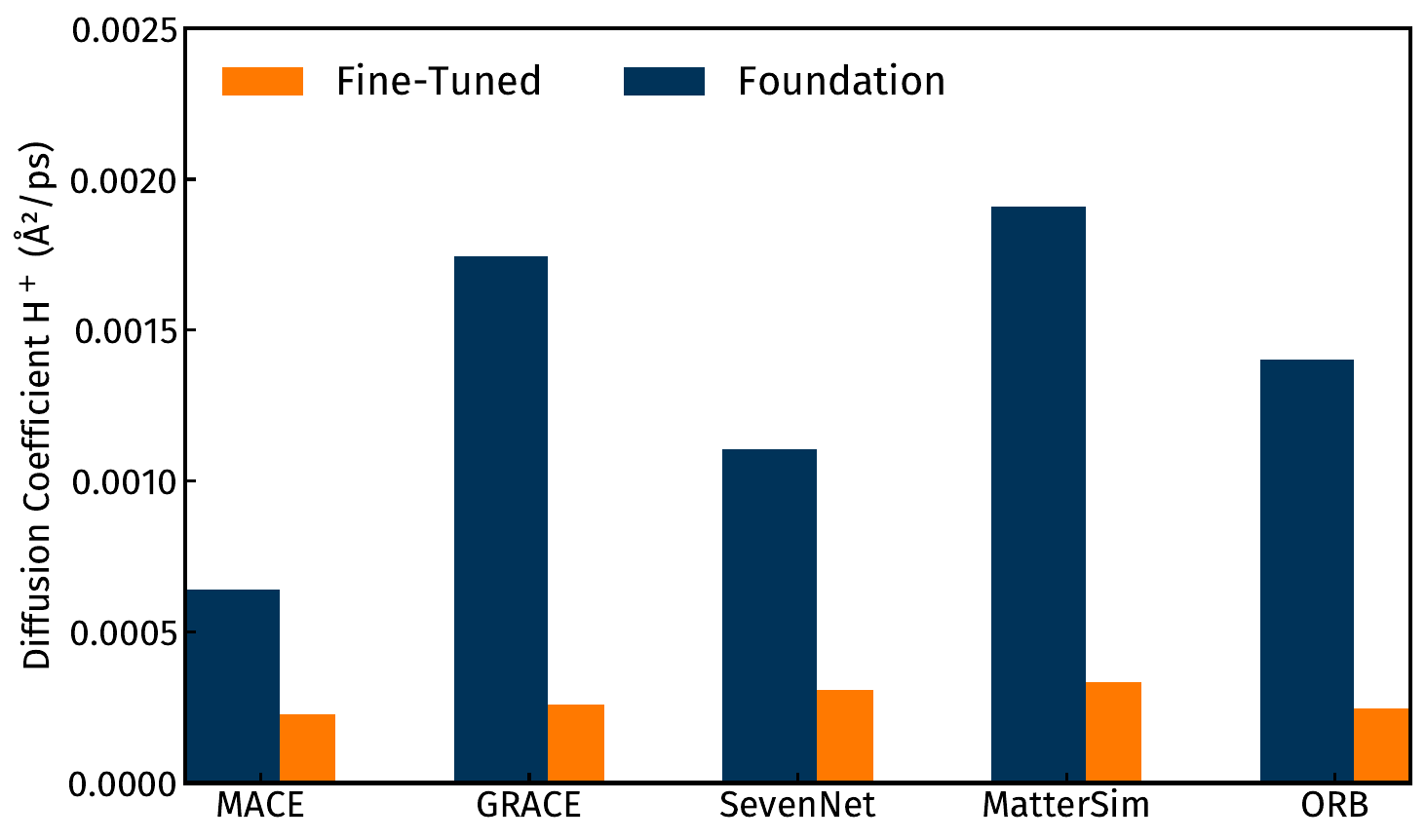}
    \caption{Diffusion coefficients of H\textsuperscript{+} in Cs\textsubscript{7}(H\textsubscript{4}PO\textsubscript{4})(H\textsubscript{2}PO\textsubscript{4})\textsubscript{8} computed 
    using different MLIP frameworks from the mean-square displacements (see Figure \ref{img:cpp_msd}). Results are shown for the foundation model and the fine-tuned foundation model. Reference AIMD data are not available, as AIMD simulations cannot provide converged diffusion coefficients for this system. For comparison, a recent MLIP study reported a diffusion coefficient of 0.0004 \AA\textsuperscript{2}/ps at 510 K.\cite{grunert2025}} 
    \label{img:cpp_diffcoeff}
\end{figure*}

\begin{figure*}[]
    \centering
    \includegraphics[width=\textwidth]{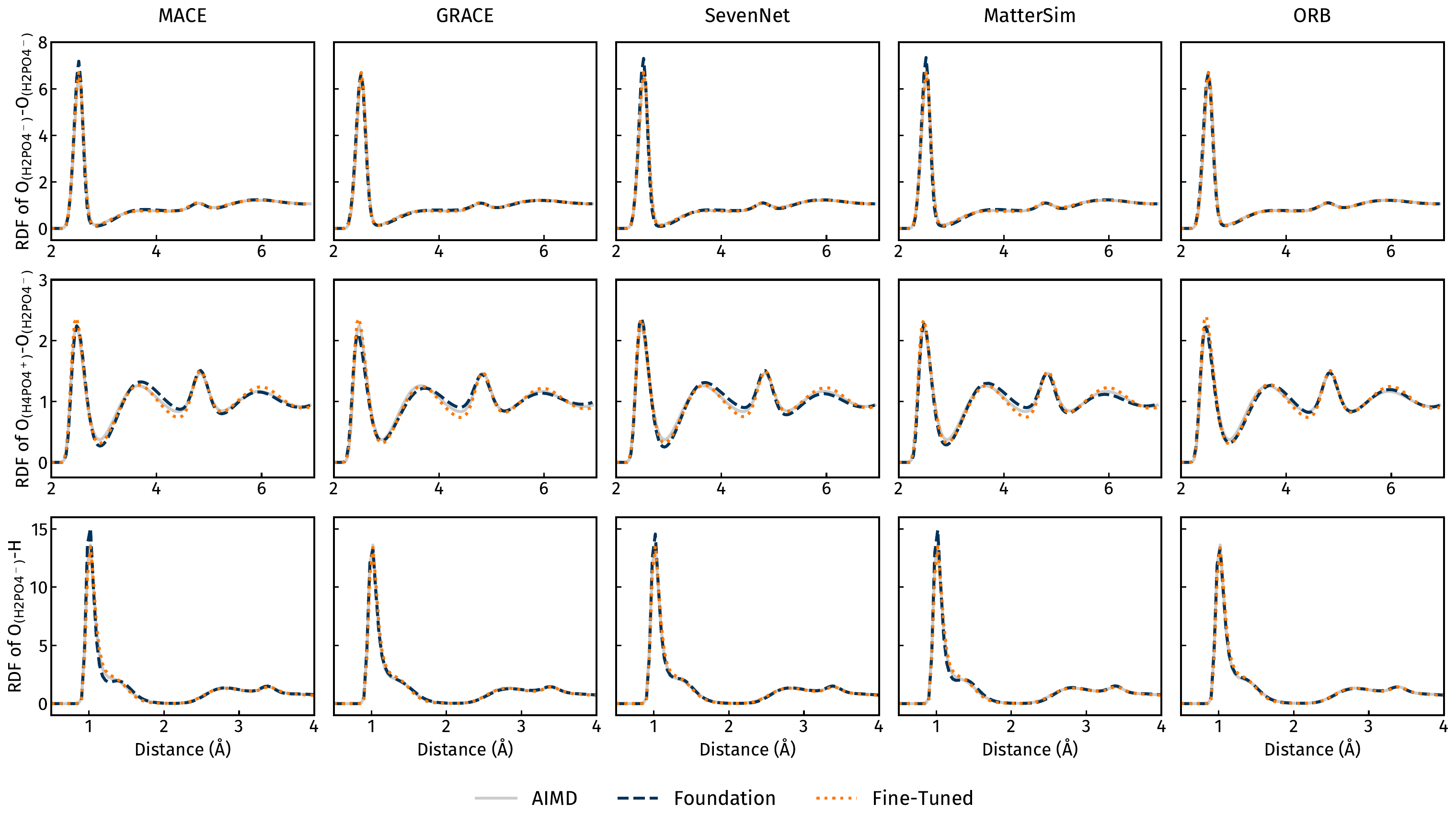}
    \caption{Radial-distribution functions of O\textsubscript{H\textsubscript{2}PO\textsubscript{4}\textsuperscript{-}}-O\textsubscript{H\textsubscript{2}PO\textsubscript{4}\textsuperscript{-}}, O\textsubscript{H\textsubscript{4}PO\textsubscript{4}\textsuperscript{+}}-O\textsubscript{H\textsubscript{2}PO\textsubscript{4}\textsuperscript{-}} and O\textsubscript{H\textsubscript{2}PO\textsubscript{4}\textsuperscript{-}}-H in Cs\textsubscript{7}(H\textsubscript{4}PO\textsubscript{4})(H\textsubscript{2}PO\textsubscript{4})\textsubscript{8} computed using different MLIP frameworks. Results from the foundation model and the fine-tuned foundation model are compared against AIMD reference data.} 
    \label{img:cpp_rdf}
\end{figure*}

\begin{figure*}[]
    \centering
    \includegraphics[width=0.75\textwidth]{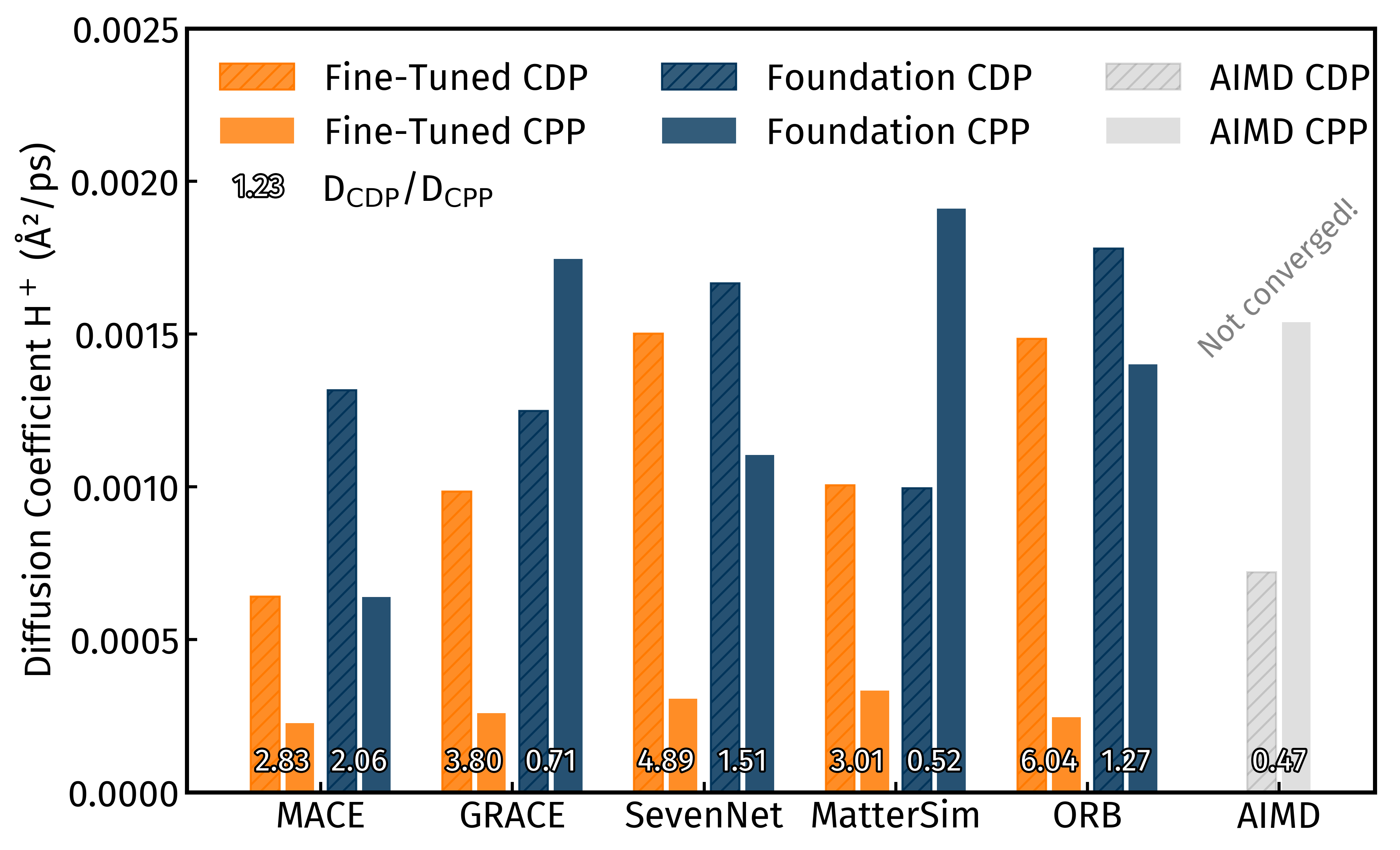}
    \caption{Diffusion coefficients comparison of H\textsuperscript{+} in CsH\textsubscript{2}PO\textsubscript{4} and Cs\textsubscript{7}(H\textsubscript{4}PO\textsubscript{4})(H\textsubscript{2}PO\textsubscript{4})\textsubscript{8} computed using different MLIP frameworks from the mean-square displacements (see Figure \ref{img:cdp_diffcoeff} and Figure \ref{img:cpp_diffcoeff}). Results are shown for the foundation model and the fine-tuned foundation model. The AIMD data result in non-converged diffusion coefficients. For comparison, a recent MLIP study reported a diffusion coefficient ratio of 4.\cite{grunert2025}} 
    \label{img:cpp_cdp_diffcoeff}
\end{figure*}

\FloatBarrier
\newpage
\section{System C: Li\textsubscript{13}Si\textsubscript{4}}

\begin{figure*}[]
    \centering
    \includegraphics[width=\textwidth]{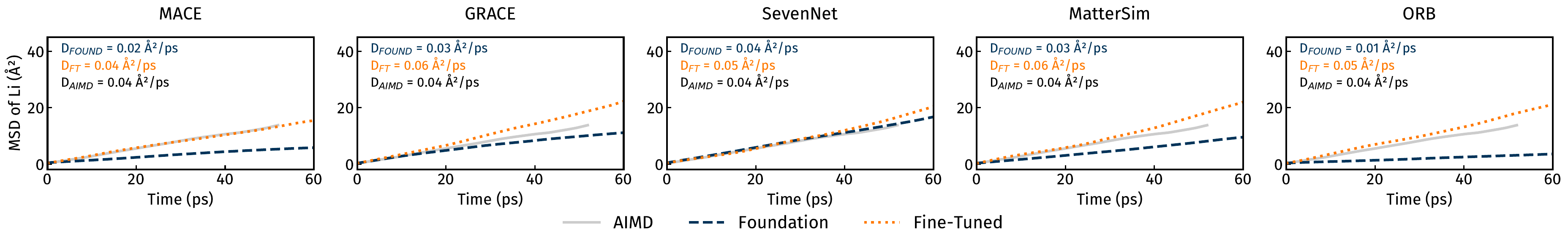}
    \caption{Mean-squared displacements of Li\textsuperscript{+} in Li\textsubscript{13}Si\textsubscript{4} computed using different MLIP frameworks. Results from the foundation model and the fine-tuned foundation model are compared against AIMD reference data.} 
    \label{img:lisi_msd}
\end{figure*}

\begin{figure*}[]
    \centering
    \includegraphics[width=0.55\textwidth]{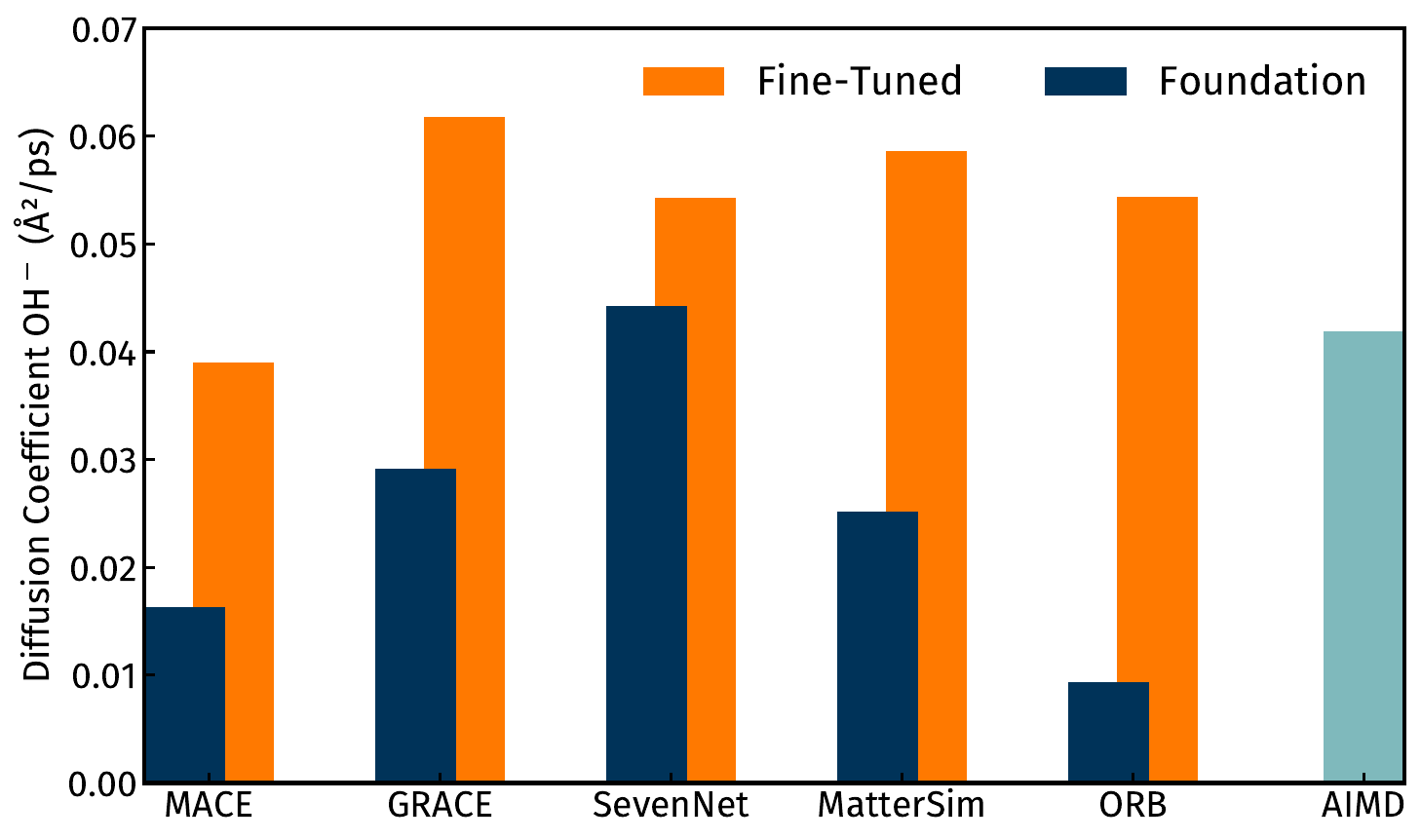}
    \caption{Diffusion coefficients of Li\textsuperscript{+} in Li\textsubscript{13}Si\textsubscript{4} computed using different MLIP frameworks from the mean-square displacements (see Figure \ref{img:lisi_msd}). Results from the foundation model and the fine-tuned foundation model are compared against AIMD reference data.} 
    \label{img:lisi_diffcoeff}
\end{figure*}

\begin{figure*}[]
    \centering
    \includegraphics[width=1\textwidth]{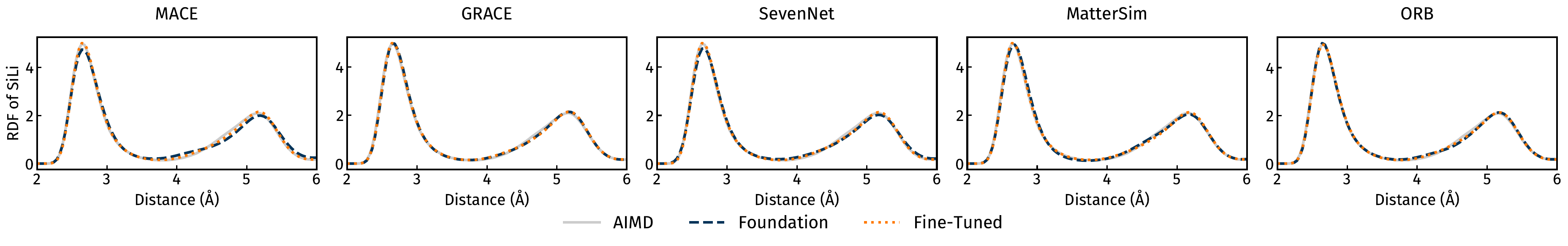}
    \caption{Radial-distribution functions of Si-Li in Li\textsubscript{13}Si\textsubscript{4} computed using different MLIP frameworks. Results from the foundation model and the fine-tuned foundation model are compared against AIMD reference data.} 
    \label{img:lisi_rdf}
\end{figure*}

\FloatBarrier
\newpage
\section{System D: PhOH in H\textsubscript{2}O}

\begin{figure*}[]
    \centering
    \includegraphics[width=\textwidth]{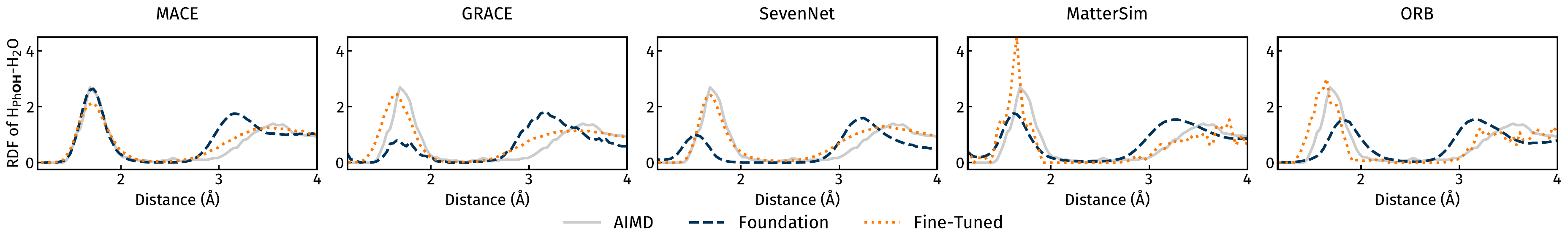}
    \caption{Radial-distribution functions of H\textsubscript{Hydroxyl-Group}-O\textsubscript{Water} in PhOH in water computed using different MLIP frameworks. Results from the foundation model and the fine-tuned foundation model are compared against AIMD reference data.} 
    \label{img:phoh_rdf}
\end{figure*}

\begin{figure*}[]
    \centering
    \includegraphics[width=\textwidth]{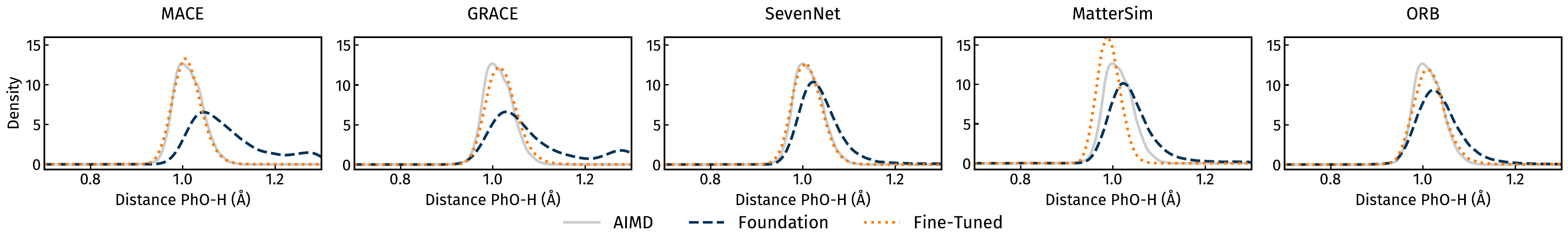}
    \caption{Distribution of the hydroxyl group O--H bond length in the phenol in water computed using different MLIP frameworks. Results from the foundation model and the fine-tuned foundation model are compared against AIMD reference data.} 
    \label{img:phoh_hdens}
\end{figure*}

\FloatBarrier
\newpage
\section{System E: KOH in H\textsubscript{2}O}

\begin{figure*}[]
    \centering
    \includegraphics[width=\textwidth]{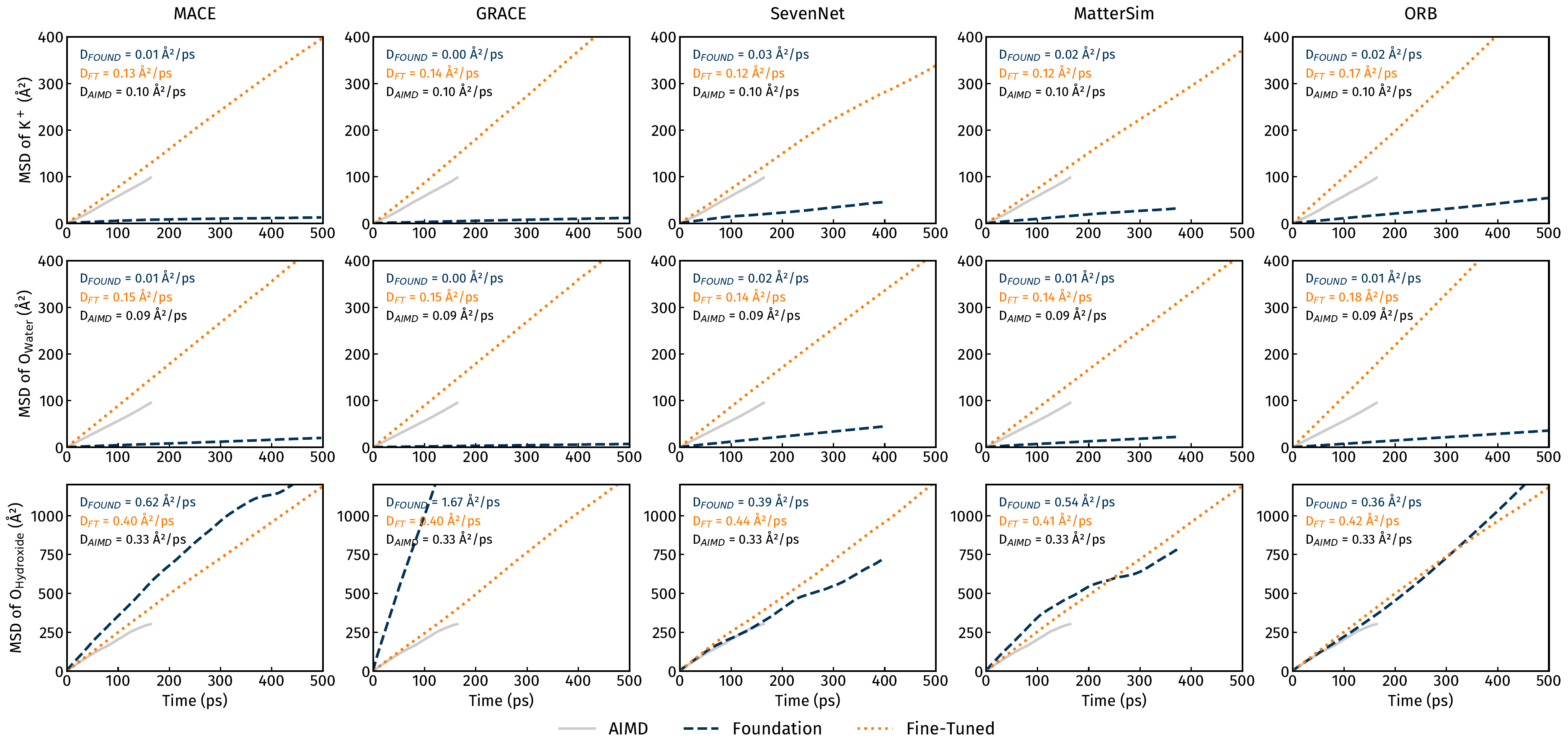}
    \caption{Mean-squared displacements of K\textsuperscript{+}, H\textsubscript{2}O and OH\textsuperscript{-} in aqueous potassium hydroxide solution computed using different MLIP frameworks. Results from the foundation model and the fine-tuned foundation model are compared against AIMD reference data.} 
    \label{img:koh_msd}
\end{figure*}

\begin{figure*}[]
    \centering
    \includegraphics[width=0.55\textwidth]{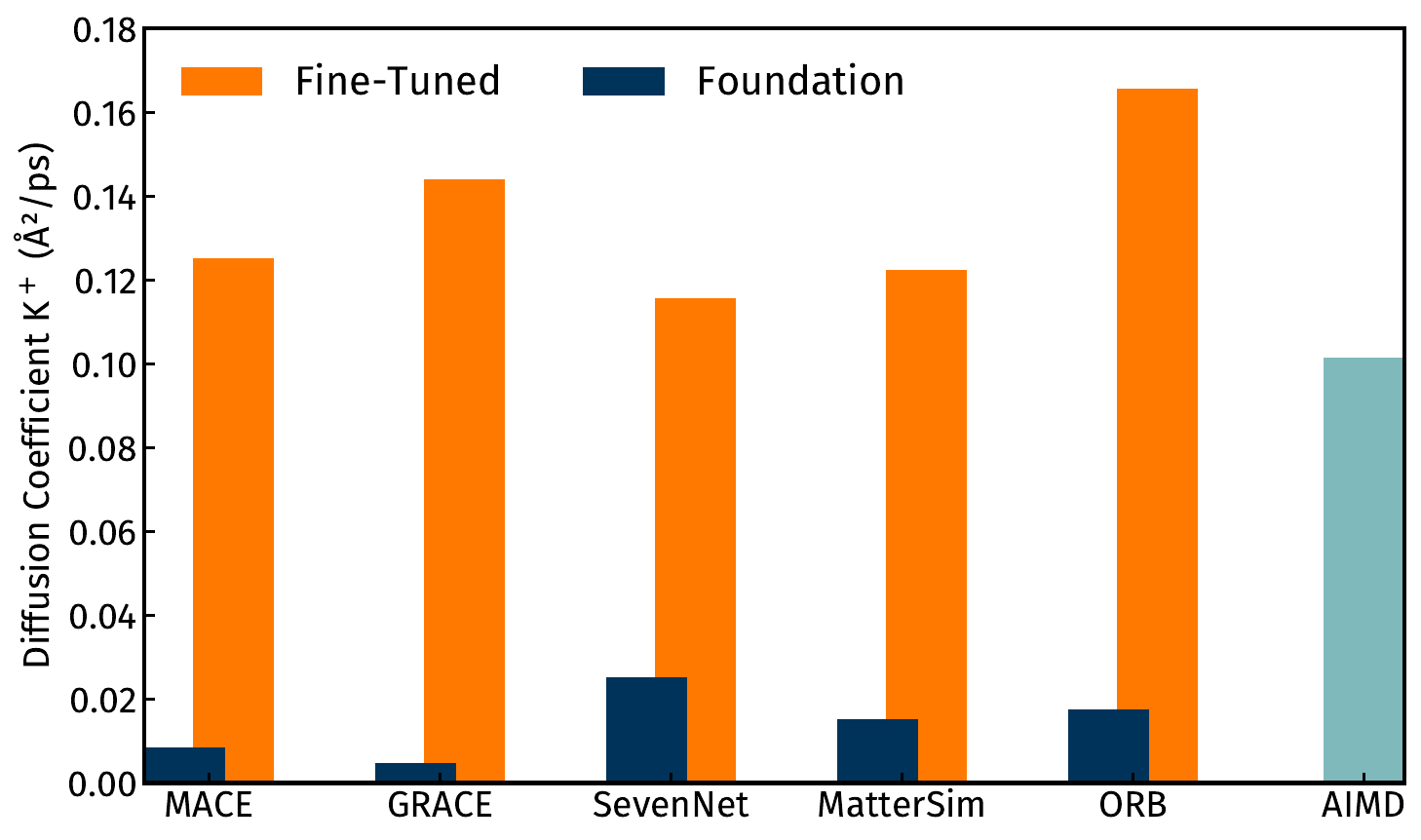}
    \caption{Diffusion coefficients of K\textsuperscript{+} in aqueous potassium hydroxide solution computed using different MLIP frameworks from the mean-square displacements (see Figure \ref{img:koh_msd}). Results from the foundation model and the fine-tuned foundation model are compared against AIMD reference data.} 
    \label{img:koh_k_diffcoeff}
\end{figure*}

\begin{figure*}[]
    \centering
    \includegraphics[width=0.55\textwidth]{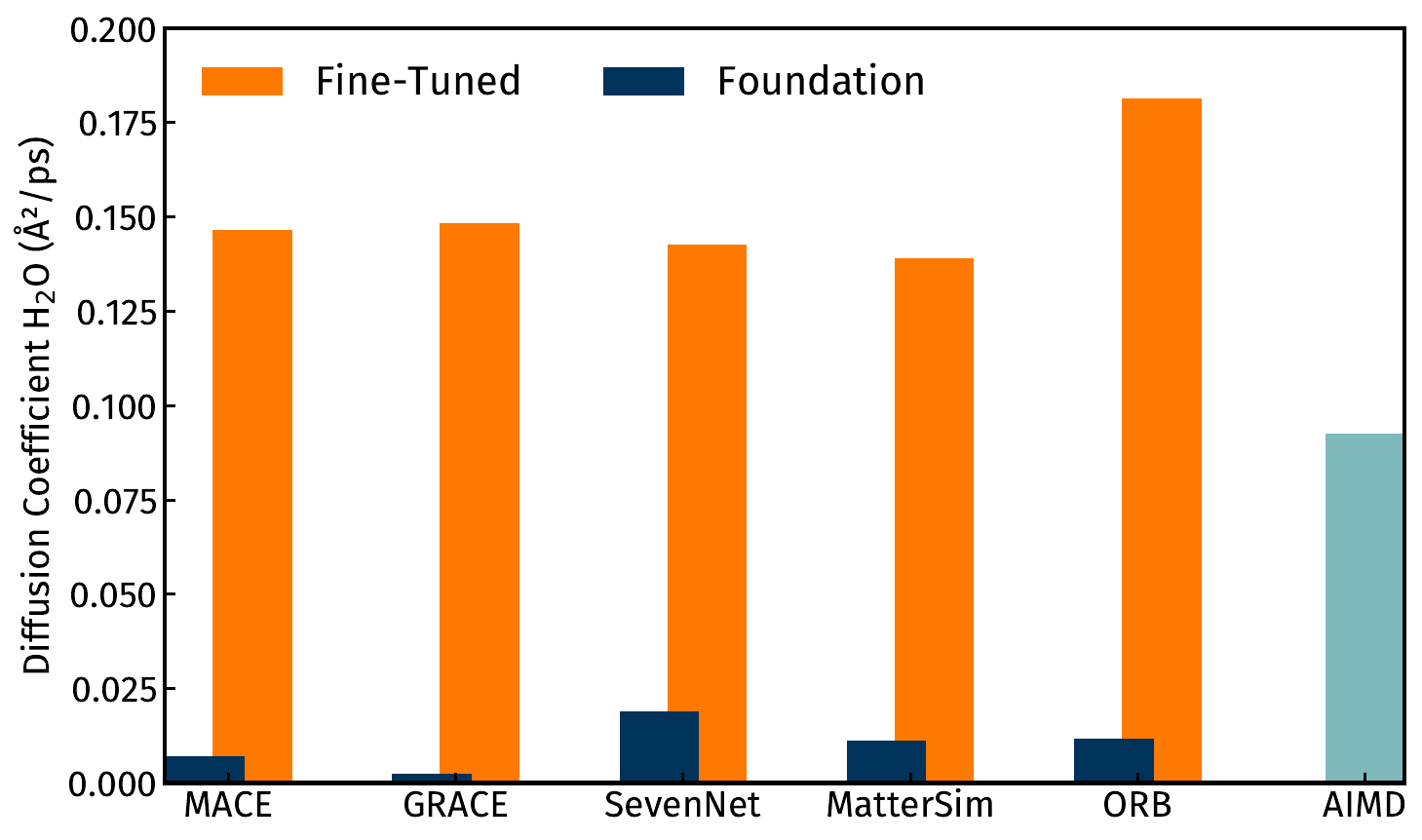}
    \caption{Diffusion coefficients of H\textsubscript{2}O in aqueous potassium hydroxide solution computed using different MLIP frameworks from the mean-square displacements (see Figure \ref{img:koh_msd}). Results from the foundation model and the fine-tuned foundation model are compared against AIMD reference data.} 
    \label{img:koh_ow_diffcoeff}
\end{figure*}

\begin{figure*}[]
    \centering
    \includegraphics[width=0.55\textwidth]{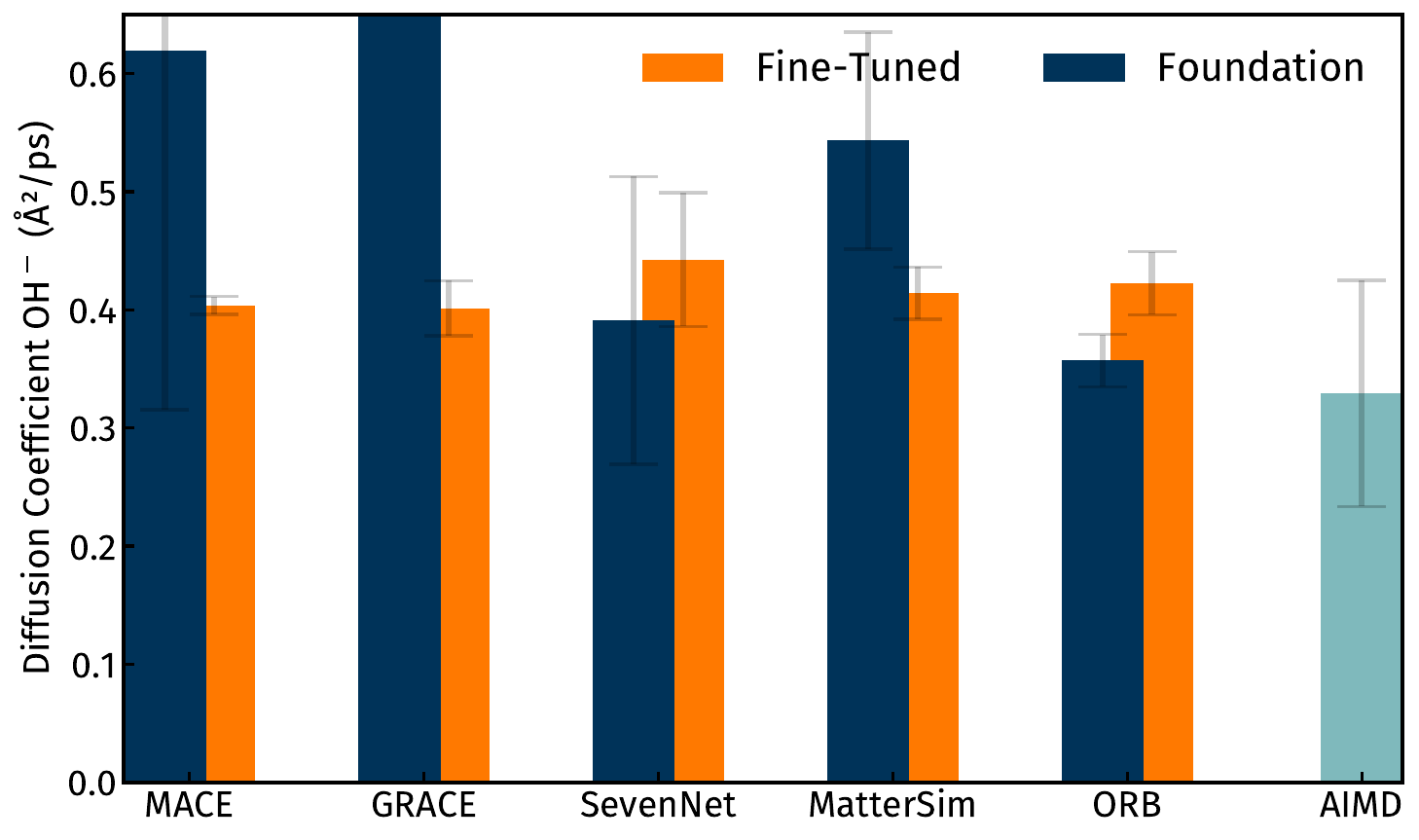}
    \caption{Diffusion coefficients of OH\textsuperscript{-} in aqueous potassium hydroxide solution computed using different MLIP frameworks from the mean-square displacements (see Figure \ref{img:koh_msd}). Results from the foundation model and the fine-tuned foundation model are compared against AIMD reference data.} 
    \label{img:koh_diffcoeff}
\end{figure*}

\begin{figure*}[]
    \centering
    \includegraphics[width=\textwidth]{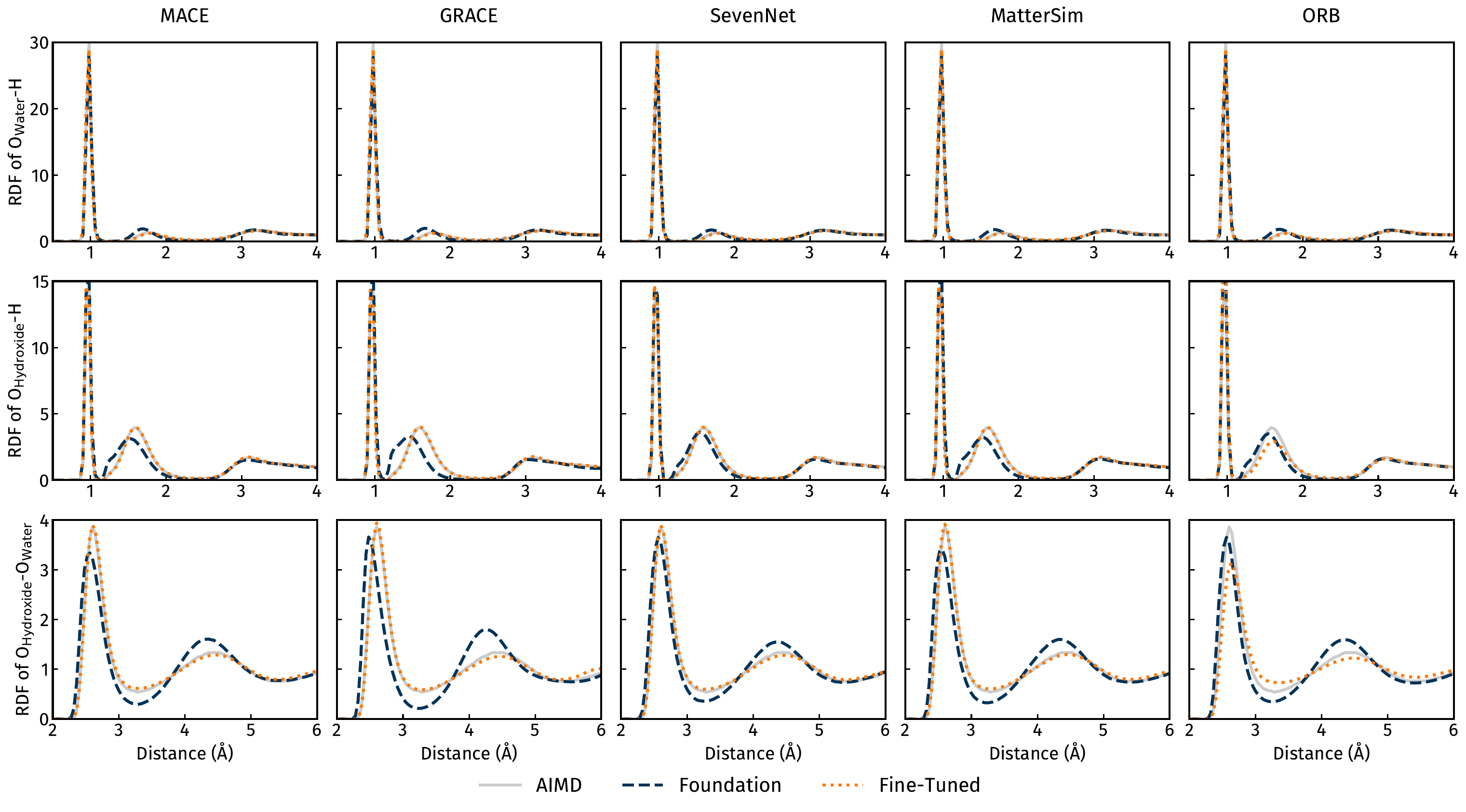}
    \caption{Radial-distribution functions of O\textsubscript{Water}-H, O\textsubscript{Hydroxide}-H and O\textsubscript{Hydroxide}-O\textsubscript{Water} in aqueous potassium hydroxide solution computed using different MLIP frameworks. Results from the foundation model and the fine-tuned foundation model are compared against AIMD reference data.} 
    \label{img:koh_rdf}
\end{figure*}

\FloatBarrier
\newpage
\section{System E: L-pyroglutamate-NH\textsubscript{4}}

\begin{figure*}[]
    \centering
    \includegraphics[width=\textwidth]{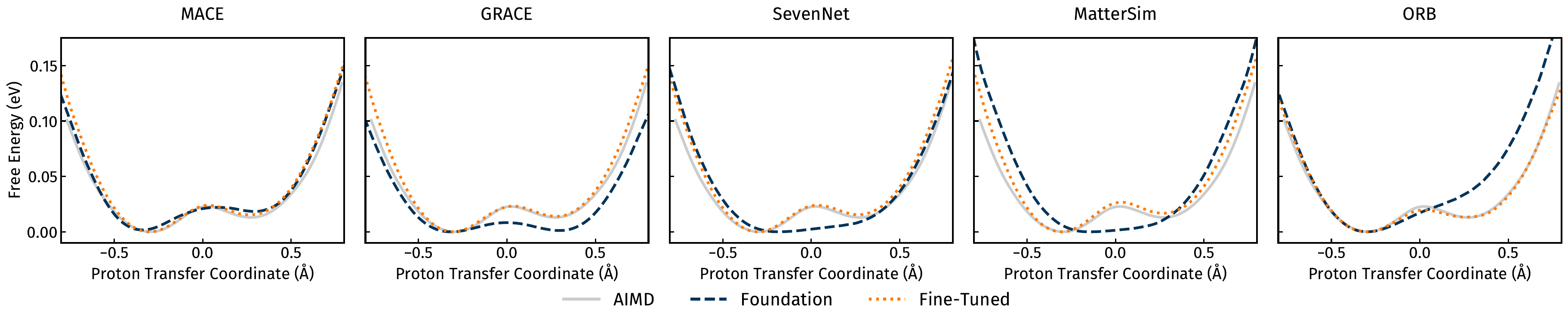}
    \caption{Free energy profiles along the proton transfer coordinate of the short-hydrogen-bond in L-pyroglutamate-NH\textsubscript{4} computed using different MLIP frameworks. Results from the foundation model and the fine-tuned foundation model are compared against AIMD reference data.} 
    \label{img:lypro_profiles}
\end{figure*}

\begin{figure*}[]
    \centering
    \includegraphics[width=\textwidth]{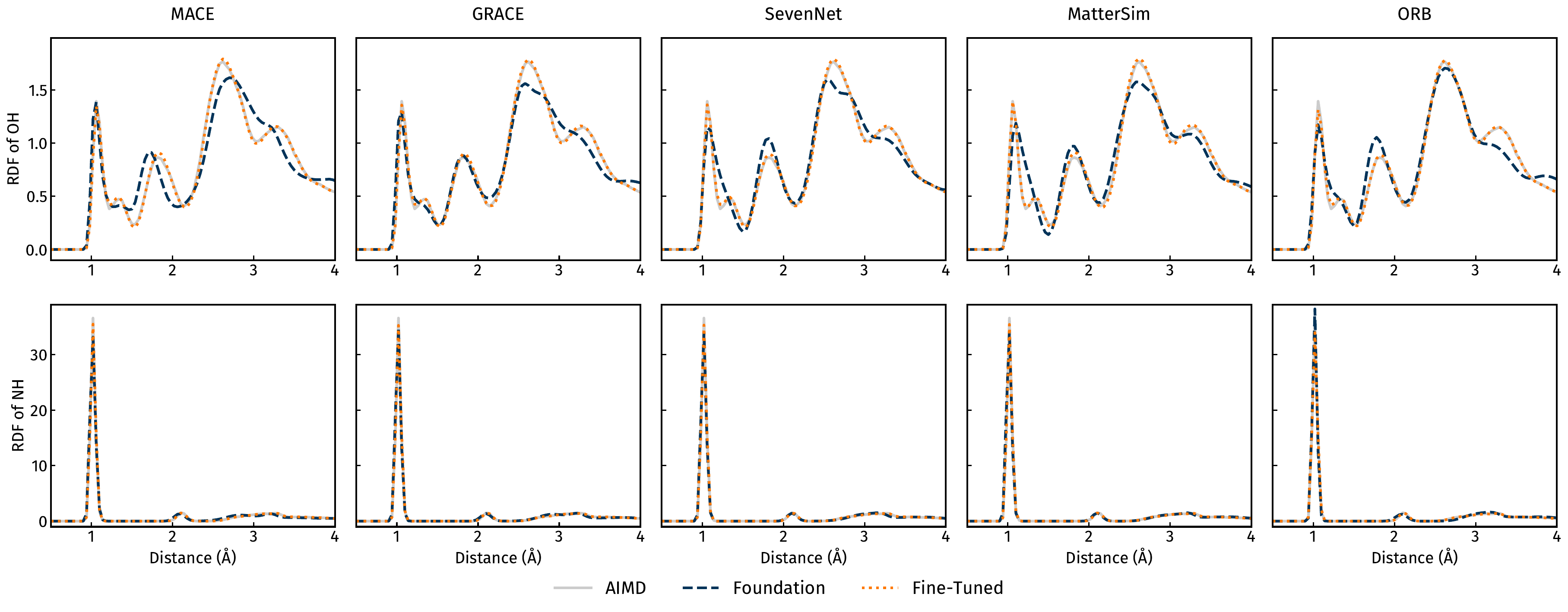}
    \caption{Radial-distribution functions of O-H and N-H in L-pyroglutamate-NH\textsubscript{4} computed using different MLIP frameworks. Results from the foundation model and the fine-tuned foundation model are compared against AIMD reference data.}
    \label{img:lpyro_rdf}
\end{figure*}

\FloatBarrier
\newpage
\section{System E: MoS\textsubscript{2}}

\begin{figure*}[]
    \centering
    \includegraphics[width=\textwidth]{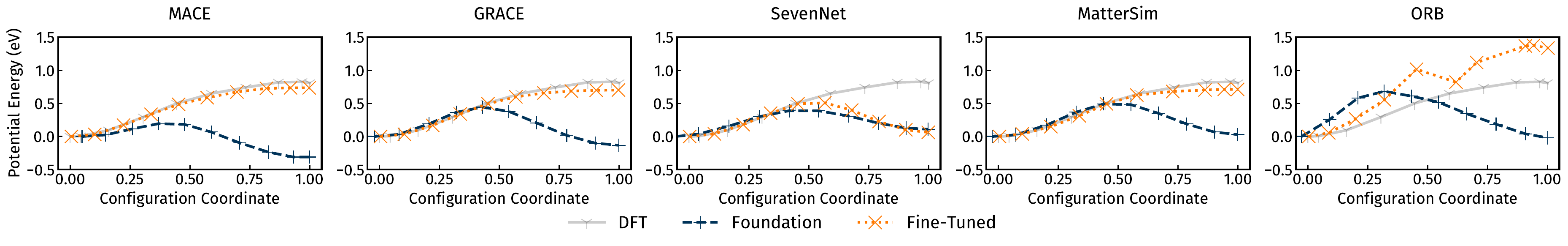}
    \caption{Potential energy curves for a sulfur jump into a sulfur vacancy cluster in MoS\textsubscript{2} computed using different MLIP frameworks. Results from the foundation model and the fine-tuned foundation model are compared against DFT reference data. Note: Fine-tuning attempts for the SevenNet foundation model did not yield models capable of reproducing the reference potential energy curve, even after hyperparameter optimization.} 
    \label{img:mos2_neb}
\end{figure*}

\FloatBarrier
\newpage

\bibliography{bibliography.bib}